\documentclass[12pt]{article}
\usepackage[utf8]{inputenc}
\usepackage{setspace}
\usepackage[linesnumbered,ruled]{algorithm2e}
\usepackage{xurl}
\usepackage{booktabs}
\usepackage{amssymb}
\usepackage{amsmath,amsthm, mathtools,commath}
\usepackage{graphics, color,epsfig}
\usepackage{subfig, subcaption}

\usepackage{multicol}
\usepackage{graphicx}
\usepackage{epsfig}
\usepackage{makeidx}
\usepackage{kotex}
\usepackage{fullpage}
\usepackage{amssymb}
\usepackage[title]{appendix}

\usepackage[utf8]{inputenc}
\usepackage[T1]{fontenc}
\usepackage{amsfonts}

\usepackage{soul}
\usepackage[super,sort&compress]{natbib}

\newcommand{\biblist}{\begin{list}{}
{\listparindent 0.0cm \leftmargin 0.50cm \itemindent -0.50 cm
\labelwidth 0 cm \labelsep 0.50 cm
\usecounter{list}}\clubpenalty4000\widowpenalty4000}
\newcommand{\ebiblist}{\end{list}}

\usepackage{latexsym}
\usepackage{amsmath, amssymb, amsfonts, amsthm, bm}
\usepackage{graphicx}
\usepackage{mathrsfs}
\usepackage{color,soul}

\newtheorem{theorem}{Theorem}

\newtheorem{proposition}{Proposition}

\newcommand{\by}{\mathbf{y}}

\newcommand{\bDelta}{{\mbox{\boldmath$\Delta$}}}
\newcommand{\bTheta}{\bm{\Theta}}
\newcommand{\btheta}{\bm{\theta}}
\newcommand{\bY}{\bm{Y}}

\makeatletter

\newcommand{\Rmnum}[1]{\expandafter\@slowromancap\romannumeral #1@}
\makeatother

\usepackage{multirow}
\usepackage{authblk}

\newcommand{\theorychange}[1]{{\color{black}#1}}
\newcommand{\notationchange}[1]{{\color{black}#1}}
\newenvironment{theoryrevision}{\par\begingroup\color{black}}{\par\endgroup}
\providecommand{\Pn}{\mathbb P_n}
\providecommand{\E}{\mathbb E}
\providecommand{\Prb}{\mathbb P}
\providecommand{\Var}{\operatorname{Var}}

\providecommand{\op}{o_p}
\providecommand{\Op}{O_p}
\providecommand{\ind}{\mathbf 1}
\providecommand{\T}{^{\mathsf T}}

\begin{document}

\baselineskip .3in
\title{Pattern Graph Models for Non-monotone Missing Data: An Application to Myocardial Infarction Complications}
\baselineskip .3in
\title{Pattern Graph Models for Non-monotone Missing Data with an Application to Myocardial Infarction Complications}

\author[1]{Yanghyeon Cho}
\author[2]{Jae Kwang Kim\thanks{Corresponding author. Department of Statistics, Iowa State University, 2438 Osborn Drive, Ames, IA 50011, USA. Email: jkim@iastate.edu. Tel: +1-515-294-3440.}}
\author[2]{Emily Berg}
\affil[1]{Department of Mathematics and Statistical Science, University of Idaho, Moscow, ID 83844, USA}

\affil[2]{Department of Statistics, Iowa State University, Ames, IA 50011, USA}

\date{}
\maketitle

\noindent\textbf{ORCID:} Yanghyeon Cho (0000-0002-3641-0162); Jae Kwang Kim (0000-0002-0246-6029); Emily Berg (0000-0002-0962-005X)
\maketitle

\begin{abstract}
Non-monotone missingness is frequently encountered in practice, with intermittent dropout and re-entry in longitudinal studies as a canonical example. Under non-monotone missingness, the missing at random (MAR) assumption often lacks a clear causal interpretation and imposes implicit constraints on the joint distribution that may be difficult to justify. Within the pattern graph framework, we propose a fully parametric estimation procedure that combines pattern-graph-identified response probabilities with a parametric outcome model through an EM algorithm with fractional imputation. \theorychange{Under correct model specification and explicit regularity conditions, we establish consistency and asymptotic normality when the fractional imputation is exact or sufficiently accurate, and, under additional moment conditions, justify parametric bootstrap variance estimation.} Simulation studies demonstrate that the proposed approach is competitive across a range of settings and provides substantial efficiency gains for joint probability estimation when fully observed cases are scarce. The methods are illustrated through an analysis of mortality risk factors in data on myocardial infarction complications.
\end{abstract}


\doublespacing
\newpage 

\section{Introduction}
\label{sec:intro}
Multivariate missing data arise routinely in biomedical and social science research \citep{schafer1998multiple, hui2004gap}. A key structural feature of multivariate missingness is its \emph{pattern}: the combination of observed and unobserved variables for each unit. The pattern is called monotone if the variables can be ordered so that whenever one variable is missing, all subsequent variables are also missing \citep{wilson2014working}. Non-monotone patterns, in which subjects may have different subsets of variables observed, are the focus of the present work. A canonical example arises in longitudinal studies with intermittent dropout and re-entry, where a subject who misses one visit may return at a later visit.

Inference with incomplete data depends critically on assumptions about the \emph{response mechanism}, the conditional distribution of the missingness pattern given the study variables. Following \citet{rubin1976inference}, the mechanism is said to be missing at random (MAR) if, conditional on the observed variables, missingness is independent of the unobserved variables; it is missing not at random (MNAR) when the mechanism also depends on unobserved values. Under monotone missingness, the MAR assumption leads to a convenient factorization of the observed-data likelihood, and estimation is relatively straightforward. Under non-monotone missingness, however, MAR is more difficult to work with because such a factorization is generally unavailable \citep{robins1997non, seaman2013}.

Several approaches have been proposed for non-monotone missing data under MAR. \citet{robins1997non} introduced the randomized monotone missingness process to produce a nontrivial MAR model. \citet{sun2018inverse} developed unconstrained maximum-likelihood and constrained Bayesian procedures to identify the MAR mechanism. \citet{kim2012factoring} extended the factoring likelihood approach by combining identified estimators from disjoint pattern-specific subsamples via generalized least squares. \citet{chen2019nonparametric} proposed nonparametric inference under donor-based identifying restrictions in pattern mixture models. 

A parallel line of work targets estimation of the response mechanism itself under non-monotone missingness. \citet{zhao2020statistical} proposed a semiparametric likelihood for the probability of complete response in which the conditional distribution of the missing variables is modeled by piecewise empirical distributions, and subsequent unified approaches employ parametric working models to accommodate general non-monotone patterns \citep{zhao2021unified, liu2022weighted, thiessen2022unified}. These methods are formulated under the MAR assumption and, for non-monotone patterns, rest on pseudo-likelihood (composite-likelihood) constructions whose large-sample theory is less complete than in the monotone case. The approach developed in this paper differs in three respects: the response mechanism is modeled directly under the MNAR structure identified by the pattern graph; estimation of the outcome model is based on the observed-data likelihood (\ref{eq:obslike}) rather than a pseudo-likelihood surrogate; and the resulting estimators are accompanied by consistency and asymptotic normality results together with a parametric bootstrap whose calibration we verify numerically.

Despite these advances, the MAR assumption in non-monotone settings has been questioned on substantive grounds. \citet{little2021missing} argued that the MAR restriction, when applied to non-monotone patterns, often lacks a clear causal interpretation: in the absence of a natural temporal ordering among variables, specifying which variables can drive missingness of which others becomes largely a modeling convenience rather than a reflection of the data-generating process. \citet{kim2021statistical} further noted that in non-monotone settings, the MAR condition imposes implicit constraints on the joint distribution that may be difficult to justify empirically. These observations suggest that MNAR models may provide a more natural framework for non-monotone missingness.

To address the above challenges, \citet{chen2022pattern} introduced the \emph{regular pattern graph} as an alternative, a tool for encoding structural assumptions about the response mechanism in non-monotone MNAR data. In this directed graph, each node represents a missingness pattern and edges encode child--parent relationships between patterns. \theorychange{The key identification assumption is a \emph{partial MAR condition}: for each non-source pattern, its probability relative to the sum of its parent-pattern probabilities depends only on variables observed under that pattern. This pooled-parent odds restriction can induce an overall MNAR mechanism.} \citet{chen2022pattern} developed three estimators under this framework: an inverse probability weighting (IPW) estimator, a regression adjustment (RA) estimator, and a semiparametrically efficient estimator based on augmented IPW (AIPW).

In this paper, we propose  a fully parametric estimation procedure based on fractional imputation (FI) \citep{kim2011parametric} under the pattern graph framework of \citet{chen2022pattern}. By specifying a parametric model $f(\bm{y}; \boldsymbol{\theta})$ for the complete data and using the pattern-graph-identified response probabilities $\pi_{\bm{\delta}}(\bm{y}; {\boldsymbol{\phi}})$, \theorychange{we first estimate $\boldsymbol{\phi}$ by conditional logistic fits, then use EM with fractional imputation to maximize the observed-data likelihood in $\boldsymbol{\theta}$ with $\hat{\boldsymbol{\phi}}$ held fixed. This yields a two-step plug-in likelihood estimator.} Unlike regression adjustment approaches that estimate pattern-specific conditional models, the proposed method performs imputation under a single joint model for the complete data. Consequently, observations from all response patterns contribute jointly to the estimation of $\bm{\theta}$ through the observed-data likelihood, allowing information to be borrowed across missingness patterns. This unified likelihood formulation mitigates the uncongeniality issues that may arise in regression adjustment approaches, as discussed in Appendix F of \citet{chen2022pattern}, and can improve estimation efficiency, particularly when fully observed cases are scarce. 

\theorychange{For a fixed graph and correctly specified odds and outcome models, we derive the influence function of this two-step estimator, including the effect of first-stage estimation on the outcome fit. We establish a consistent sandwich estimator and justify parametric bootstrap variance estimation under local uniformity and moment conditions. Exact categorical enumeration requires no Monte Carlo approximation; continuous-data FI requires additional control of imputation bias and variability.} Extensive simulation studies support the theoretical results and demonstrate substantial efficiency gains for joint probability estimation when the proportion of fully observed cases is low. The proposed method is illustrated with an analysis of the associations of hypotension and opioid administration with mortality among patients with myocardial infarction.

The remainder of the paper is organized as follows. Section~\ref{sec:setup} reviews the pattern graph framework and the estimators of \citet{chen2022pattern}. Section~\ref{sec:proposed} presents the parametric fractional imputation method. Section~\ref{sec:theory} establishes the asymptotic theory. Section~\ref{sec:simulation} reports simulation studies, and Section~\ref{sec:application} illustrates the methods with data on myocardial infarction complications. Section~\ref{sec:conclusion} concludes with a discussion.

\section{Review of the Pattern Graph Approach}\label{sec:setup}
In this section, we first introduce the setup and notation used
throughout the paper, and then review the pattern graph approach of
\citet{chen2022pattern}, on which our proposed method builds.
\subsection{Pattern graph models for response probabilities}\label{sec:patterngraph}
Let $\bm{y}_{1},\hdots,\bm{y}_{n}$ be independently and identically distributed (i.i.d.) realizations of a random vector $\bm{Y}\!=\!(Y_{1},\hdots,Y_{p})^\top \in \Omega \subset \mathbb{R}^p$. Each component of $\bm{Y}$ can be missing, producing  non-monotone missing patterns. Let $\bm{\Delta}= (\Delta_{1},\hdots,\Delta_{p})^\top \in 
 \{0,1\}^p$ denote the response pattern, where $\Delta_k=1$ if $Y_k$ is observed and 0 otherwise.   
We use 
$\mathcal{D}\subset \{0,1\}^p$ to denote the set of response patterns that occur with positive probability. We assume that $\bm{1} \in \mathcal{D}$.  
Let $\bm{\delta}_1, \hdots, \bm{\delta}_n$ be the $n$ realizations of $\bm{\Delta}$. For $\bm{\delta}\in\mathcal{D}$, let 
$S(\bm{\delta})=\{k:\delta_k=1\}$ and 
$S^c(\bm{\delta})=\{k:\delta_k=0\}$ 
denote the sets of observed and unobserved indices, respectively. 
We write $\bm{Y}_{S(\bm{\delta})}$ and $\bm{Y}_{S^c(\bm{\delta})}$ 
for the corresponding observed and unobserved subvectors of $\bm{Y}$. 
For example, when $p=3$ and $\bm{\delta}=(1,0,1)$, 
we have $\bm{Y}_{S(\bm{\delta})}=(Y_1,Y_3)^\top$ and 
$\bm{Y}_{S^c(\bm{\delta})}=(Y_2)$. 
We are interested in estimating $\zeta_{0}=\mathbb{E}[h(\bm{Y})]$ for a known function $h$.

Following 
\citet{chen2022pattern}, we use 
 a regular pattern graph on the node set $\mathcal{D}$ to encode the identification restrictions in non-monotone missing data. \begin{theoryrevision}
Each node $\bm\delta\in\mathcal D$ represents a missingness pattern,
and an edge $\bm u\to\bm v$ identifies $\bm u$ as a parent of $\bm v$.
Write $\pi_{\bm\delta}(\bm y)=P(\bm\Delta=\bm\delta\mid\bm Y=\bm y)$.
The odds model compares each non-source pattern with its pooled parent
set, as specified below.
\end{theoryrevision}

 A regular pattern graph satisfies the following two conditions:
\begin{itemize}
    \item[(i)] The graph has a unique source, a pattern with no parent, given by the complete pattern $\bm{1}=(1,\hdots, 1)$.
    \item[(ii)] For any edge $\bm{u}\rightarrow\bm{v}$, $\bm{u} > \bm{v}$, 
where $\bm{u} > \bm{v}$ denotes coordinate-wise dominance with at least one strict inequality.
\end{itemize}

Let $P_{\bm{\delta}}$ denote the set of parents of $\bm{\delta} \in \mathcal{D}$. Intuitively, parents contain strictly more observed components than their children, enabling identification ``from above''. We make the following assumptions: 

\begin{enumerate}
\item 
\begin{theoryrevision}
{\bf Positivity}: For each non-source pattern
$\bm\delta\in\mathcal D\setminus\{\bm 1\}$ and almost every $\bm g$ in the
relevant support of $\bm Y_{S(\bm\delta)}$, assume
\[
P(\bm\Delta=\bm\delta\mid\bm Y_{S(\bm\delta)}=\bm g)>0,\qquad
P(\bm\Delta\in P_{\bm\delta}\mid\bm Y_{S(\bm\delta)}=\bm g)>0.
\]
Also assume $\pi_{\bm 1}(\bm y)>0$ almost everywhere on the outcome
support. Parent-set positivity is not imposed on the source, whose
parent set is empty.
\end{theoryrevision}

\item
\begin{theoryrevision}
{\bf Identification via partial MAR odds.}
Let $\pi_{P_{\bm\delta}}(\bm y)=
\sum_{\bm\kappa\in P_{\bm\delta}}\pi_{\bm\kappa}(\bm y)$.
For each $\bm\delta\in\mathcal D\setminus\{\bm 1\}$, assume
\end{theoryrevision}

\begin{align}\label{FI_cond_odds}
\frac{\pi_{\bm{\delta}}(\bm{y})}{\pi_{P_{\bm{\delta}}}(\bm{y})}
=\frac{\mathbb{P}(\bm{\Delta}=\bm{\delta}\mid \bm{Y}_{S(\bm{\delta})}= 
\bm{y}_{S(\bm{\delta})})}{\mathbb{P}(\bm{\Delta}\in P_{\bm{\delta}}\mid
\bm{Y}_{S(\bm{\delta})}= 
\bm{y}_{S(\bm{\delta})})}:=O_{\bm{\delta}}\{ 
\bm{y}_{S(\bm{\delta})}\}.
\end{align} 
\end{enumerate}

This partial MAR condition concerns only the ratio of probabilities between a pattern and its parents, conditional on the variables observed under 
$\bm{\delta}$. The overall mechanism $\{ \pi_{\bm{\delta}} (\by) \}$ may still depend on unobserved components and thus can be MNAR.

The partial MAR condition in (\ref{FI_cond_odds}) makes it possible to identify the response probability $\pi_{\bm{\delta}}(\bm{y})$ from the observed data. Define the relative odds from the source as $C_{\bm{\delta}}(\bm{y})={\pi_{\bm{\delta}}(\bm{y})}/{\pi_{\bm{1}}(\bm{y})}$. From (\ref{FI_cond_odds}), we can express 
\begin{align}\label{FI_cond_source}
C_{\bm{\delta}}(\bm{y})
=\frac{\pi_{\bm{\delta}}(\bm{y})}{\pi_{P_{\bm{\delta}}}(\bm{y})}\frac{\pi_{P_{\bm{\delta}}}(\bm{y})}{\pi_{{\bm{1}}}(\bm{y})}= O_{\bm{\delta}} \{ 
\bm{y}_{S(\bm{\delta})}\} 
\sum_{\bm{\kappa}\in P_{\bm{\delta}}}C_{\bm{\kappa}}(\bm{y}). 
\end{align}
By topologically sorting the regular pattern graph starting from the unique source $\bm{1}= (1, \ldots, 1)$, setting $C_{{\bm1}}\equiv 1$, and applying (\ref{FI_cond_source}) down the graph, all $C_{\bm{\delta}}( \by)$ are identified. Since $\sum_{\bm{\delta} \in \mathcal{D}} \pi_{\bm \delta} ( \by ) = 1$, we obtain 
$\pi_{\bm{1}}(\bm{y}) =\{ \sum_{\bm{\delta}\in\mathcal{D}}C_{\bm{\delta}}(\bm{y})\}^{-1}$,
and the response probability for  $\bm{\Delta} = \bm{\delta}$ is given by $\pi_{\bm\delta}(\bm{y}) = \pi_{\bm{1}}(\bm{y})C_{\bm{\delta}}(\bm{y}).$  
The recursion in (\ref{FI_cond_source}) thus identifies $\pi_{\bm{\delta}}( \by)$ from these MAR-type odds. 

In practice, we model the conditional odds via a logistic regression on the observed components under $\bm{\delta}$: 
\begin{equation}
  O_{\bm{\delta}}(\bm{y}_{S(\bm{\delta})};\bm{\phi}_{\bm{\delta}})=\exp\big((1,\bm{y}_{S(\bm{\delta})}^\top )\bm{\phi}_{\bm{\delta}}\big),  \label{logistic}
\end{equation}
so that 
$ \mathbb{P} \left( \bDelta = \bm{\delta} \mid \bDelta \in \{ \bm{\delta} \} \cup P_{\bm{\delta}} , \bm{Y}_{S(\bm{\delta})} = \bm{y}_{S(\bm{\delta})} ; \bm{\phi}_{\bm{\delta}} \right)=  \mbox{expit}\big((1,\bm{y}_{S(\bm{\delta})}^\top )\bm{\phi}_{\bm{\delta}}\big).$
We estimate the model parameter in (\ref{logistic}) by maximizing the binary log-likelihood over units with $\bm{\Delta} \in \{ \bm{\delta} \} \cup P_{\bm{\delta}}$: 
\begin{equation}
\hat{\bm{\phi}}_{\bm{\delta}} = \mbox{arg} \max_{\bm{\phi}_{\bm{\delta}}} \sum_{i=1}^n  \left[\mathbb{I}(\bm{\Delta}_{i}=\bm{\delta}) \log \left\{ p_{i \bm{\delta}} (\bm{\phi}_{\bm{\delta}}) \right\}   + \mathbb{I}(\bm{\Delta}_{i}\in P_{\bm{\delta}})
\log \left\{ 1-  p_{i \bm{\delta}} (\bm{\phi}_{\bm{\delta}}) \right\}\right]\label{mle}
\end{equation}
\begin{theoryrevision}
where, for eligible units in $\{\bm\delta\}\cup P_{\bm\delta}$,
\[
\begin{aligned}
p_{i\bm\delta}(\bm\phi_{\bm\delta})
&=P(\bm\Delta_i=\bm\delta\mid
\bm\Delta_i\in\{\bm\delta\}\cup P_{\bm\delta},
\bm Y_{i,S(\bm\delta)}=\bm y_{i,S(\bm\delta)};\bm\phi_{\bm\delta})\\
&=\operatorname{expit}\{(1,\bm y_{i,S(\bm\delta)}^\top)
\bm\phi_{\bm\delta}\}.
\end{aligned}
\]
Only the coordinates observed under the comparison pattern $\bm\delta$
enter this regression. Contributions from ineligible units are zero.
\end{theoryrevision}

\theorychange{With $\hat{\bm{\phi}}=\{\hat{\bm{\phi}}_{\bm{\delta}}:\bm{\delta}\in\mathcal D\setminus\{\bm 1\}\}$, compute} 
$ \hat{O}_{\bm{\delta}} \left\{  \bm{Y}_{S(\bm{\delta})}  \right\}= {O}_{\bm{\delta}} \{\bm{Y}_{S(\bm{\delta})} ; \hat{\bm{\phi}}_{\bm{\delta}} \}$ in (\ref{logistic}) and evaluate the recursion  \eqref{FI_cond_source} in topological order, setting 
 $C_{\bm{1}}(\bm{Y})=1$. We can compute $\hat{\pi}_{\bm{\delta}} (\bm{Y})$ as follows: 
\begin{enumerate}
\item \theorychange{For each non-source $\bm\delta\in\mathcal D\setminus\{\bm 1\}$, use~\eqref{mle} to obtain $\hat{\bm\phi}_{\bm\delta}$ and $\hat O_{\bm\delta}$.} 
    \item Compute $\hat{C}_{\bm{\delta}}(\bm{Y})=\hat{O}_{{\bm{\delta}}}(\bm{Y}_{S({\bm{\delta}})})\sum_{\bm{\kappa}\in P_{\bm{\delta}}}\hat{C}_{\bm{\kappa}}(\bm{Y})$ 
starting from the source. 
    
    \item Obtain $\hat{\pi}_{\bm{1}} \left( \bm{Y} \right)= \left\{ \sum_{\bm{\delta} \in \mathcal{D}} \hat{C}_{\bm{\delta}} (\bm{Y} ) \right\}^{-1}$ and compute 
    $ \hat{\pi}_{\bm{\delta}}(\bm{Y} )  = \hat{\pi}_{\bm{1}} \left( \bm{Y} \right) \cdot
\hat{C}_{\bm{\delta}} (\bm{Y} ) .     
    $
    \end{enumerate}

\begin{figure}
\centering
\subfloat[]{\includegraphics[width=0.0425 \textwidth]{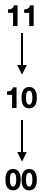}\label{FI_4figs-a}}
\hfill
\subfloat[]{ \includegraphics[width=0.20\textwidth]{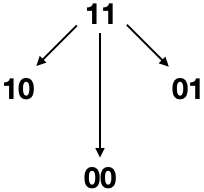}\label{FI_4figs-b}}%
\hfill
\subfloat[]{ \includegraphics[width=0.20\textwidth]{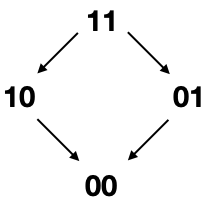}\label{FI_4figs-c}}%
\hfill
\subfloat[]{ \includegraphics[width=0.20\textwidth]{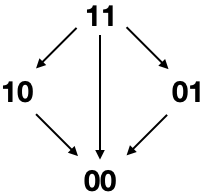}\label{FI_4figs-d}}%
\caption{Possible regular pattern graphs when a bivariate variable is subject to missingness.}
\label{FI_4figs}
\end{figure}

For example, given the pattern graph in Figure \ref{FI_4figs} (c), there are four nodes in $\mathcal{D} = \{ 11, 10, 01, 00\}$ with edges $ 11 \rightarrow 10$, $11 \rightarrow 01$, and $\{10, 01\} \rightarrow 00$. Then we can compute $\hat{\pi}_{\bm{\delta}}(\bm{Y})=\hat{\pi}_{11}(\bm{Y})\hat{C}_{\bm{\delta}}(\bm{Y}),$ where 
    $ \hat{C}_{10} = \hat{O}_{10} (Y_1), \ \ \hat{C}_{01} = \hat{O}_{01} (Y_2), \ \ \hat{C}_{00} = \hat{O}_{00} ( \hat{C}_{10} + \hat{C}_{01} )
    $
and 
    $$\hat{\pi}_{11}(\bm{Y})=
    \left( 1+ \hat{C}_{10} + \hat{C}_{01} + \hat{C}_{00} \right)^{-1} = \left[ 1+ \left( 1+ \hat{O}_{00} \right) \left\{ \hat{O}_{10} (Y_1) + \hat{O}_{01} (Y_2) 
    \right\} 
    \right]^{-1}.$$
These $\hat{\pi}_{\bm{\delta}} ( \bm{Y})$ will be used in Section \ref{sec:proposed} to construct the proposed estimator. 

\subsection{Existing methods under the pattern graph framework}\label{sec:reviewofchen}

Under the pattern graph model of \citet{chen2022pattern}, one can construct a direct estimator based only on the completely observed units. This estimator is called the inverse probability weighting (IPW) estimator, as each complete unit is weighted by the inverse of the probability of having a full response ${\pi}_{\bm{1}}(\bm{y})$. In our setup, the IPW estimator of $\zeta$ is   defined as the solution to
\begin{align}\label{FI_IPW}
\sum_{i=1}^{n}\frac{\mathbb{I}(\bm{\delta}_{i}=\bm{1})}{\hat{\pi}_{\bm{1}}(\bm{y}_{i})} \left\{h(\bm{y}_{i}) - \zeta \right\} =0,
\end{align}
where $\hat{\pi}_{\bm{1}}(\bm{y})$ is the estimated pattern probability for $\bm{\delta}=\bm{1}$, obtained
using the procedure described in Section \ref{sec:patterngraph}.

Another approach is the regression adjustment (RA) estimator, which can be represented as 
\begin{align}\label{FI_RA}
    \hat{\zeta}_{\rm RA} = \frac{1}{n} \sum_{i=1}^{n}E[h(\bm{Y}_{i})\mid\by_{S(\bm\delta_i)},\bm{\delta}_{i}].
\end{align} 
To compute the conditional expectation in (\ref{FI_RA}) with non-monotone missing data, \citet{chen2022pattern} proposed a novel sequential regression imputation under the regular pattern graph model.

Lastly, we can combine the two estimators to  construct an  augmented IPW (AIPW) estimator, which consists of the IPW estimator and an augmentation term. The latter is an element in the augmentation space of \citet{tsiatis2006semiparametric}, defined as 
$\Lambda=\{g(\bm{\Delta},\bm{Y}_{S(\bm{\Delta})}): E\big(g(\bm{\Delta},\bm{Y}_{S(\bm{\Delta})}\mid\bm{Y})\big)=0\}$. 
Thus, the AIPW estimator maintains unbiasedness of the IPW estimator while improving its efficiency by incorporating the incomplete units through the augmentation term. The key step in the AIPW method is selecting the most efficient estimator from the class of unbiased estimators for $\zeta_{0}$ defined as: $$\bigg\{\frac{h(\bm{Y})\mathbb{I}(\bm{\Delta}=\bm{1})}{\hat{\pi}_{\bm{1}}(\bm{Y})}+g(\bm{\Delta},\bm{Y}_{S(\bm{\Delta})}): E\big(g(\bm{\Delta},\bm{Y}_{S(\bm{\Delta})})\mid\bm{Y}\big)=0\bigg\}.$$

Given a pattern graph, define $\Gamma_{\bm{s},\bm{r}}$ as the set of all paths from $\bm{s}$ to $\bm{r}$.
\citet{chen2022pattern} proposed the following AIPW estimator: 
\begin{align}\label{FI_SEMI}
   \hat{\zeta}_{\rm AIPW} = \frac{1}{n} \sum_{i=1}^{n}\bigg[\frac{h(\bm{Y}_{i})\mathbb{I}(\bm{\Delta}_{i}=\bm{1})}{\hat{\pi}_{\bm{1}}(\bm{Y}_{i})}+\sum_{\bm{r}\neq\bm{1}}\sum_{\bm{s}\in A_{\bm{r}}}{\widehat{\rm EIF}}_{\bm{s},\bm{r}}(\bm{Y}_{i},\bm{\Delta}_{i})\bigg],
\end{align}
where ${\widehat{\rm EIF}}_{\bm{s},\bm{r}}(\bm{Y}_{i},\bm{\Delta}_{i})=\hat{M}_{i,\bm{s}}\big[\mathbb{I}(\bm{\Delta}_{i}=\bm{s})-\hat{O}_{\bm{s}}(\bm{Y}_{i,S(\bm{s})})\mathbb{I}(\bm{\Delta}_{i}\in P_{\bm{s}})\big]\sum_{\gamma \in \Gamma_{\bm{s},\bm{r}}}\prod_{\bm{t}\in\gamma,\bm{t}<\bm{s}}\hat{O}_{\bm{t}}(\bm{Y}_{i,S(\bm{t})})$, $A_{\bm{r}}=P_{\bm{r}}\bigcup\{\bm{r}\}$, $\hat{M}_{i,\bm{s}}= \hat{E}[h(\bm{Y}_{i})|\bm{Y}_{i,S(\bm{s})},\bm{\Delta}_i=\bm{s}]$. Here, $\bm{Y}_{i,S(\bm{\delta})} = (Y_{ik}: k\in S(\bm{\delta}))$ denotes the subvector of $\bm{Y}_i$ corresponding to the observed indices under pattern $\bm{\delta}$.

For instance, under the pattern graph in Figure \ref{FI_4figs} (c), the AIPW estimator can be obtained by:
\begin{equation}\label{FI_SEMI_chen_1c}
\begin{aligned}
      \hat{\zeta}_{\rm AIPW} 
         &=\frac{1}{n}\sum_{i=1}^{n}\bigg[\frac{\mathbb{I}(\bm{\Delta}_{i}=11)h(\bm{Y}_{i})}{\hat{\pi}_{\bm{1}}(\bm{Y}_{i})}\\
      &+[\mathbb{I}(\bm{\Delta}_{i}=10)-\hat{O}_{10}(Y_{1i}) \mathbb{I}(\bm{\Delta}_{i}=11)]\,\hat{M}_{i,10}\big(1+\hat{O}_{00}\big)\\
      &+[\mathbb{I}(\bm{\Delta}_{i}=01)-\hat{O}_{01}(Y_{2i}) \mathbb{I}(\bm{\Delta}_{i}=11)]
      \,\hat{M}_{i,01}\big(1+\hat{O}_{00}\big)\\
      &+[\mathbb{I} (\bm{\Delta}_{i}=00)-\hat{O}_{00}\,{\mathbb{I} (\bm{\Delta}_{i} \in \{ 10, 01 \} } )] \,\hat{M}_{i,00}\bigg],
\end{aligned}
\end{equation}
where the conditional expectation $E[h(\bm{Y}_{i})\mid\bm{Y}_{i,S(\bm{r})},\bm{r}]$ can be estimated using the RA method.

\section{Proposed Method Using a Parametric Outcome Model}\label{sec:proposed}

\subsection{Estimation procedure}

We assume a parametric outcome model with density \(f(\by;\bm{\theta})\), \(\bm{\theta}\in\bm{\Theta}\). The joint density of \((\bY,\bDelta)\) is modeled as
\begin{equation}
f(\by,{\bm{\delta}};\bm{\eta})\;=\;\Pr(\bDelta={\bm{\delta}}\mid \by;\bm{\phi})\,f(\by;\btheta),\qquad \bm{\eta}=(\bm{\phi},\btheta)\in\bm{\Phi}\times\bTheta,
\label{FI_Full_function}
\end{equation}
where \(\pi_{\bm{\delta}}(\by;\bm{\phi}):=\Pr(\bDelta={\bm{\delta}}\mid \by;\bm{\phi})\) is identified via the pattern-graph odds in Section~\ref{sec:setup}. \theorychange{With missing data, the observed-data likelihood is}
\begin{equation}
L_{\mathrm{obs}}(\bm{\phi},\btheta)\;=\;\prod_{i=1}^n \int f(\by_i,{\bm{\delta}}_i;\bm{\eta})\,d\by_{S^{c}(\bm{\delta}_i)}
\;=\;\prod_{i=1}^n \int \pi_{\bm{\delta}_i}(\by;\bm{\phi})\,f(\by;\btheta)\,d\by_{S^{c}(\bm{\delta}_i)}.
\label{eq:obslike}
\end{equation}
\begin{theoryrevision}
The integral in~\eqref{eq:obslike} couples the odds and outcome
parameters. We use a sequential two-step procedure. First,
$\bm\phi$ is estimated by the conditional logistic likelihood~\eqref{mle}
for each non-source pattern against its pooled parent set. This step
uses no outcome model, and the resulting estimator is consistent under
correct odds specification, identification, and regularity conditions,
even if the outcome model is misspecified.
Second, $\bm\theta$ maximizes
$L_{\rm obs}(\hat{\bm\phi},\bm\theta)$ with $\hat{\bm\phi}$ held fixed.
This two-step plug-in likelihood estimator can be less efficient than
the joint maximum likelihood estimator when both models are correctly specified. Its influence
function~\eqref{eq:IF_FI} contains the term
\[
\{d_{\bm\phi}^{\top}
-d_{\bm\theta}^{\top}A_{\bm\theta}^{-1}C\}
A_{\bm\phi}^{-1}u_{\bm\phi},
\]
which accounts for the direct effect of $\hat{\bm\phi}$ on imputation
and its indirect effect through $\hat{\bm\theta}$. These quantities are
defined in Section~\ref{sec:theory}. Both stages are re-estimated in
every bootstrap replicate.
\end{theoryrevision}

Let \(\ell_{\mathrm{com}}(\bm{\phi},\btheta)
= n^{-1}\sum_{i=1}^n\{\log \pi_{\bm{\delta}_i}(\bY_i;\bm{\phi})+\log f(\bY_i;\btheta)\}\).
The EM iteration computes
\begin{equation}
Q(\btheta\,|\,\btheta^{(t)})\;=\;\mathbb{E}\!\left[\ell_{\mathrm{com}}(\hat{\bm{\phi}},\btheta)\,\middle|\,\mathcal{O}_{\mathrm{obs}};\,\btheta^{(t)}\right],
\label{eq:7} 
\end{equation}
\theorychange{where $\mathcal O_{\rm obs}=\{O_{i,\rm obs}=(\bm\delta_i,\bm y_{i,S(\bm\delta_i)}):i=1,\ldots,n\}$}
and sets \(\btheta^{(t+1)}=\arg\max_{\btheta} Q(\btheta\,|\,\btheta^{(t)})\).
Because \(\bm{\phi}=\hat{\bm{\phi}}\) is fixed, the \(\log\pi_{\bm{\delta}_i}\) term is constant in \(\bm{\theta}\); the M-step reduces to weighted maximum likelihood estimation of \(f(\cdot;\bm{\theta})\).

The central computational challenge is evaluating the conditional expectation in the E-step (\ref{eq:7})  under the current model. When this integral lacks a closed-form expression, as is typically the case for non-monotone patterns with multiple missing variables, we approximate it using parametric fractional imputation \citep{kim2011parametric}. The idea is straightforward: for each incomplete unit $i$, draw multiple candidate values $\bm{y}^{*(j)}_{S^c(\bm{\delta}_i)}$ for the missing part from a proposal distribution $h_m (\cdot)$ that has broad support, and assign each candidate a fractional weight proportional to the ratio of its density under the current model to the proposal density. The weighted average over these candidates then approximates the conditional expectation. As the model parameter $\boldsymbol{\theta}$ is updated across EM iterations, the fractional weights are recalculated while the imputed values remain fixed, making the procedure computationally efficient.

Specifically, 
we evaluate \eqref{eq:7} using the \emph{parametric fractional imputation} of \citet{kim2011parametric}. For each incomplete unit \(i\), draw \(m_i\) proposals for the missing part only,
$\by_{S^{c}(\bm{\delta}_i)}^{*(j)}\sim h_m(\cdot\mid \by_{S(\bm{\delta}_i)},\bm{\delta}_i)$,
where \(h_m\) has full support wherever \(\pi_{\bm{\delta}_i}(\by;\hat{\bm{\phi}})f(\by;\btheta^{(t)})\) is positive; then form the completed vectors \(\by_i^{*(j)}=(\by_{S(\bm\delta_i)},\,\by_{S^{c}(\bm\delta_i)}^{*(j)})\).
With \(\btheta^{(t)}\) fixed, assign fractional weights
\[
 w^*_{ij} ( \btheta^{(t)}) \ \propto\
\frac{\pi_{\bm\delta_i}(\by_i^{*(j)};\hat{\bm{\phi}})\, f(\by_i^{*(j)};\btheta^{(t)})}
{h_m\!\left(  \by_{S^{c}(\bm\delta_i)}^{*(j)} 
\mid \by_{S(\bm\delta_i)},\bm{\delta}_i\right)},
\qquad
\sum_{j=1}^{m_i} w^*_{ij} ( \btheta^{(t)})=1.
\]
Then
\[
\theorychange{Q(\btheta\,|\,\btheta^{(t)}) \;\approx\; \frac{1}{n}\sum_{i=1}^n\sum_{j=1}^{m_i} w^*_{ij} ( \btheta^{(t)})\,\log f(\by_i^{*(j)};\btheta)\;+\;\text{const},}
\]
and the M-step maximizes the weighted log-likelihood in \(\btheta\).
Finally, the estimator of \(\zeta_0=\mathbb{E}\{h(\bY)\}\) is
\begin{equation}
\hat\zeta_{\mathrm{FI}}=\frac{1}{n}\sum_{i=1}^n\sum_{j=1}^{m_i}w^*_{ij}(\hat\btheta)\,h (\by_i^{*(j)} ),
\qquad \sum_{j=1}^{m_i}w^*_{ij}(\hat\btheta)=1.
\label{FI_FI}
\end{equation}
\begin{theoryrevision}
With exact conditional expectations, consistency of
\eqref{FI_FI} follows under correct specification and identification of
both the odds and outcome models, uniform convergence, and appropriate
optimization. Consistency of $\hat{\bm\phi}$ alone is insufficient.
For categorical outcomes, enumeration of all missing-value combinations
evaluates the E-step exactly: weights are the normalized target masses
$\pi_{\bm\delta_i}(\bm y_i^{*(j)};\hat{\bm\phi})
f(\bm y_i^{*(j)};\bm\theta)$, without division by a proposal density.
For Monte Carlo FI, proposal support,
importance-weight moments, and a sufficiently increasing imputation
count must be controlled as in Section~\ref{sec:theory} and Appendix C of Supporting Information.
Convergence of EM to an arbitrary stationary point does not ensure the
required solution.
\end{theoryrevision}

\subsection{Variance estimation}\label{FI_var}

The variability in (\ref{FI_FI}) comes from multiple sources, including the sampling scheme, the response mechanism, and the imputation. The combination of these sources complicates the derivation of an analytical variance formula, and resampling methods, such as the jackknife or the bootstrap, can be used as an alternative. To assess the accuracy of the proposed estimator (\ref{FI_FI}), we estimate its variance using the bootstrap approach.

The bootstrap procedure of \citet{efron1994missing}, called the full-mechanism bootstrap, can be applied to our missing data problems. In the full-mechanism bootstrap, the whole process can be summarized in the following three steps: (1) The complete data are generated by random sampling from either a finite or infinite population. (2) Then, the response mechanism is applied, creating the missing data. (3) Lastly, the parameter of interest ${\zeta}$ is estimated with the observed data. This perspective exactly matches our setup, as we factorize the joint density as $f(\bm{y},\bm{\delta};\bm{\eta})=P(\bm{\Delta}=\bm{\delta}\mid\bm{Y}=\bm{y};\bm{\phi})f(\bm{y};\bm{\theta})$. Next, this process is repeated multiple times, generating the bootstrap analog $\hat{\zeta}^{*\notationchange{(b)}}$ of the estimator $\hat{\zeta}$. One of the main differences from \citet{efron1994missing} is that we use a fully parametric bootstrap, since the complete data are directly modeled through the model parameter $\bm{\theta}$ and the response mechanism is also parametrically modeled through $\bm{\phi}$. 
\begin{enumerate}
    \item Generate a bootstrap sample $\{\bm{y}_{i}^{*{\notationchange{(b)}}}:i=1,\hdots,n\}$ from $f(\cdot\,;\hat{\bm{\theta}})$.
    \item Draw \({\bm{\delta}}_i^{*\notationchange{(b)}}\sim \mathrm{Mult}\!\left(\{\hat{\pi}_{\bm\delta}(\by_i^{*\notationchange{(b)}};\hat{\bm{\phi}})\}_{\bm\delta\in \mathcal{D}}\right)\), where $\sum_{\bm{\delta} \in \mathcal{D}} 
\hat{\pi}_{\bm\delta}(\by_i^{*\notationchange{(b)}};\hat{\bm{\phi}}) =1   $.  Define the bootstrap version of the observed data set, denoted as $\mathcal{O}_{\rm obs}^{*\notationchange{(b)}}=\{O_{i,\rm obs}^{*\notationchange{(b)}}=(\by_{\notationchange{i},S({\bm{\delta}_i^{*\notationchange{(b)}}})}^{*(b)},\bm{\delta}_i^{*\notationchange{(b)}})    
    ;i=1,\hdots,n\}. $ 
    \item Re-estimate \(\notationchange{\hat{\bm{\phi}}}^{*\notationchange{(b)}}\) from \(\mathcal{O}^{*\notationchange{(b)}}_{\mathrm{obs}}\) using the logistic fits of Section~\ref{sec:setup}; then re-estimate \(\notationchange{\hat{\btheta}^{*(b)}}\) by running the EM/FI with \(\notationchange{\hat{\bm{\phi}}^{*(b)}}\) fixed. Compute \(\hat\zeta^{*(b)}_{\mathrm{FI}}\).    
    
    \item Repeat Steps 1--3 $B$ times for a large $B$. Then, calculate the bootstrap estimator of $\mathrm{Var}(\hat{\zeta}_{\rm FI})$ given by $\widehat{V}_{\rm BT}=(B-1)^{-1}\sum_{b=1}^{B}(\hat{\zeta}_{\rm FI}^{*(b)}-\bar{\zeta}_{\rm FI}^{*})^{2}$, $\theorychange{\bar{\zeta}_{\rm FI}^{*}=B^{-1}\sum_{b=1}^{B}\hat{\zeta}_{\rm FI}^{*(b)}}$.
\end{enumerate}

\begin{theoryrevision}
Theorem~\ref{thm:FI_bootstrap} justifies
this variance estimator under correct specification and explicit local
uniformity and moment conditions, with $B=B_n\to\infty$ for asymptotic
consistency of the simulated variance. A fixed $B$ retains bootstrap
Monte Carlo error. The centered bootstrap distribution uses
$\zeta(\hat{\bm\theta})=\int h(\bm y)f(\bm y;\hat{\bm\theta})\,d\bm y$;
this choice of center does not affect the sample variance above. For continuous
data, refitting the same fixed-$m$ algorithm does not by itself remove
imputation bias or establish valid coverage.
\end{theoryrevision}

\begin{theoryrevision}
\noindent\textit{Remark.} The theory in Section~\ref{sec:theory}
uses i.i.d.\ observations. A survey-weighted analogue would use inverse
inclusion probabilities $d_i=1/P(I_i=1)$, where $I_i$ is the sampling indicator for unit $i$, and, for example, the normalized mean
\[
\frac{\sum_{i=1}^n d_i\sum_j w_{ij}^*h(\bm y_i^{*(j)})}{\sum_{i=1}^n d_i}.
\]
Such an extension also requires design-appropriate fitting and variance
estimation; it is not covered by the i.i.d.\ parametric bootstrap theorem.
For resampling under survey designs with missing data, see
\citet{shao1996bootstrap}, \citet{shao2003impact}, and \citet{mashreghi2016survey}.
\end{theoryrevision}


\begin{theoryrevision}
\section{Theoretical Properties}\label{sec:theory}

We study the two-step estimator for a fixed regular pattern graph, fixed
regression bases, and finite-dimensional parameters. The observations
$O_i=({\bm{\Delta}}_i,{\bm{Y}}_{i,S({\bm{\Delta}}_i)})$ are i.i.d. Let
${\bm{\eta}}=({\bm{\phi}}\T,{\bm{\theta}}\T)\T$, with true value ${\bm{\eta}}_0$, and write
\begin{equation}
g_{\bm{\eta}}(o)=\int\pi_{\bm{\delta}}({\bm{y}};{\bm{\phi}})f({\bm{y}};{\bm{\theta}})\,dy_{S^c({\bm{\delta}})},
\qquad o=({\bm{\delta}},{\bm{y}}_{S({\bm{\delta}})}).
\label{eq:theory_g}
\end{equation}
Define $m(o;{\bm{\eta}})=\E_{\bm{\eta}}\{h({\bm{Y}})\mid O=o\}$,
$\zeta({\bm{\theta}})=\E_{\bm{\theta}} h({\bm{Y}})$, and
$\Pn a=n^{-1}\sum_{i=1}^n a(O_i)$. We first consider exact evaluation
of conditional expectations, for which
\begin{equation}
\begin{aligned}
\widehat{\bm{\phi}}_{\bm{\delta}}&\in\arg\max_{{\bm{\phi}}_{\bm{\delta}}}\Pn\ell_{\bm{\delta}}(O;{\bm{\phi}}_{\bm{\delta}}),\\
\widehat{\bm{\theta}}&\in\arg\max_{\bm{\theta}}\Pn\log g_{\widehat{\bm{\phi}},{\bm{\theta}}}(O),\\
\widehat\zeta&=\Pn m(O;\widehat{\bm{\eta}}).
\end{aligned}
\label{eq:two_step_exact}
\end{equation}
Here $\ell_{\bm{\delta}}$ is the conditional logistic log-likelihood contribution
in~\eqref{mle}, defined only for non-source patterns. Exact categorical
enumeration gives $\widehat\zeta=\hat\zeta_{\rm FI}$ in~\eqref{FI_FI}.
The numerical approximation requirements are stated separately below.

\subsection{Estimating functions and the influence function}

For ${\bm{\delta}}\ne\mathbf 1$, set
$T_{\bm{\delta}}=\ind\{{\bm{\Delta}}\in\{{\bm{\delta}}\}\cup P_{\bm{\delta}}\}$,
$Z_{\bm{\delta}}=\ind\{{\bm{\Delta}}={\bm{\delta}}\}$,
${\bm{x}}_{\bm{\delta}}=(1,{\bm{Y}}_{S({\bm{\delta}})}\T)\T$, and
$p_{\bm{\delta}}=\operatorname{expit}({\bm{x}}_{\bm{\delta}}\T{\bm{\phi}}_{\bm{\delta}})$.
The first-stage score contribution is
\begin{equation}
u_{{\bm{\phi}},{\bm{\delta}}}(O;{\bm{\phi}}_{\bm{\delta}})
={\bm{x}}_{\bm{\delta}}(Z_{\bm{\delta}}-T_{\bm{\delta}} p_{\bm{\delta}}).
\label{eq:theory_uphi}
\end{equation}
It is defined to be zero when $T_{\bm{\delta}}=0$, so it uses only observed
covariates. Stack these contributions as $u_{\bm{\phi}}$.
The second-stage score, with ${\bm{\phi}}$ held fixed, is
\begin{equation}
u_{\bm{\theta}}(O;{\bm{\eta}})=\nabla_{\bm{\theta}}\log g_{\bm{\eta}}(O)
=\E_{\bm{\eta}}\{s_{\bm{\theta}}({\bm{Y}};{\bm{\theta}})\mid O\},
\qquad s_{\bm{\theta}}({\bm{y}};{\bm{\theta}})=\nabla_{\bm{\theta}}\log f({\bm{y}};{\bm{\theta}}).
\label{eq:theory_utheta}
\end{equation}
This is a plug-in observed-data score, not a profile score obtained by
reoptimizing ${\bm{\phi}}$. Moreover, $u_{\bm{\phi}}$ is generally different from the
${\bm{\phi}}$ score of $g_{\bm{\eta}}$. Under correct specification both scores have
mean zero.

All expectations in the following definitions use the fixed true law:
\begin{align}
A_{\bm{\phi}}&=-\E_0\frac{\partial u_{\bm{\phi}}}{\partial{\bm{\phi}}\T}
=\operatorname{blockdiag}_{{\bm{\delta}}\ne\mathbf 1}
\E_0\{T_{\bm{\delta}} p_{\bm{\delta}}(1-p_{\bm{\delta}}){\bm{x}}_{\bm{\delta}} {\bm{x}}_{\bm{\delta}}\T\},
\label{eq:theory_Aphi}\\
A_{\bm{\theta}}&=-\E_0\frac{\partial u_{\bm{\theta}}}{\partial{\bm{\theta}}\T},
\qquad C=-\E_0\frac{\partial u_{\bm{\theta}}}{\partial{\bm{\phi}}\T},
\label{eq:theory_AthetaC}\\
d_{\bm{\phi}}&=\E_0\{\nabla_{\bm{\phi}} m(O;{\bm{\eta}}_0)\},
\qquad d_{\bm{\theta}}=\E_0\{\nabla_{\bm{\theta}} m(O;{\bm{\eta}}_0)\}.
\label{eq:theory_d}
\end{align}
Derivatives are evaluated at ${\bm{\eta}}_0$. In particular, $d_{\bm{\phi}},d_{\bm{\theta}}$
are nonrandom averages of conditional-mean derivatives; the derivative
does not act on the outer probability law.

\begin{proposition}[Consistency]\label{prop:eta_consistency}
Under Conditions (A1)--(A3) and the consistency part of (A6) in Appendix A of Supporting Information,
$\widehat{\bm{\phi}}\xrightarrow{p}{\bm{\phi}}_0$,
$\widehat{\bm{\theta}}\xrightarrow{p}{\bm{\theta}}_0$, and
$\widehat\zeta\xrightarrow{p}\zeta_0=\zeta({\bm{\theta}}_0)$.
These conditions include correct specification of both models,
identification, uniform convergence of the criteria, and asymptotic
maximization. Consistency of $\widehat{\bm{\phi}}$ alone does not establish the
last two conclusions.
\end{proposition}

\begin{theorem}[Asymptotic linearity and normality]\label{thm:FI_normality}
Under Conditions (A1)--(A6) in Appendix A of Supporting Information, the nuisance estimators have
influence functions
\begin{align}
a_{\bm{\phi}}(O)&=A_{\bm{\phi}}^{-1}u_{\bm{\phi}}(O;{\bm{\phi}}_0),\label{eq:theory_IFphi}\\
a_{\bm{\theta}}(O)&=A_{\bm{\theta}}^{-1}\{u_{\bm{\theta}}(O;{\bm{\eta}}_0)-C A_{\bm{\phi}}^{-1}u_{\bm{\phi}}(O;{\bm{\phi}}_0)\}.
\label{eq:theory_IFtheta}
\end{align}
Moreover,
\begin{equation}
\begin{aligned}
\sqrt n(\widehat\zeta-\zeta_0)
&=n^{-1/2}\sum_{i=1}^n\psi_{\rm FI}(O_i)+\op(1)
\xrightarrow{d}N(0,V_{\rm FI}),
\\
V_{\rm FI}&=\E_0\{\psi_{\rm FI}(O)^2\}.
\end{aligned}
\label{eq:V_FI}
\end{equation}
where
\begin{equation}
\begin{aligned}
\psi_{\rm FI}(O)={}&m(O;{\bm{\eta}}_0)-\zeta_0
+d_{\bm{\theta}}\T A_{\bm{\theta}}^{-1}u_{\bm{\theta}}(O;{\bm{\eta}}_0)\\
&+\{d_{\bm{\phi}}\T-d_{\bm{\theta}}\T A_{\bm{\theta}}^{-1}C\}
A_{\bm{\phi}}^{-1}u_{\bm{\phi}}(O;{\bm{\phi}}_0).
\end{aligned}
\label{eq:IF_FI}
\end{equation}
\end{theorem}

The cross-derivative $C$ accounts for the effect of estimating ${\bm{\phi}}$ on
the outcome fit. Both this indirect effect and the direct derivative
$d_{\bm{\phi}}$ are required. Write $d=(d_{\bm{\phi}}\T,d_{\bm{\theta}}\T)\T$.
The stacked score $U=(u_{\bm{\phi}}\T,u_{\bm{\theta}}\T)\T$
has negative expected derivative
$A=\left(\begin{smallmatrix}A_{\bm{\phi}}&0\\C&A_{\bm{\theta}}\end{smallmatrix}\right)$,
whose inverse gives~\eqref{eq:IF_FI}. Conditional logistic fits may share
observations; their full covariance must therefore be retained, even
though $A_{\bm{\phi}}$ is block diagonal. Appendix B of Supporting Information gives the proofs and
conditional-covariance formulas for computing the derivatives.
No joint-likelihood or semiparametric efficiency claim is implied.

\subsection{Sandwich and bootstrap variance estimation}

Replace the expectations in~\eqref{eq:theory_Aphi}--\eqref{eq:theory_d}
by empirical averages at $\widehat{\bm{\eta}}$, and evaluate~\eqref{eq:IF_FI}
at these estimates to obtain $\widehat\psi_i$, using $\widehat\zeta$ for
$\zeta_0$. Under the local square-integrability conditions in (A4),
\begin{equation}
\widehat V_{\rm FI}=\frac1n\sum_{i=1}^n
(\widehat\psi_i-\overline{\widehat\psi})^2\xrightarrow{p}V_{\rm FI},
\qquad
\widehat{\Var}(\widehat\zeta)=\widehat V_{\rm FI}/n.
\label{eq:FI_sandwich}
\end{equation}
For $V_{\rm FI}>0$, this gives a first-order valid Wald interval.

\begin{theorem}[Parametric bootstrap]\label{thm:FI_bootstrap}
Generate ${\bm{Y}}_i^*\sim f(\cdot;\widehat{\bm{\theta}})$ and
${\bm{\Delta}}_i^*\mid {\bm{Y}}_i^*\sim\pi_{\bm{\delta}}({\bm{Y}}_i^*;\widehat{\bm{\phi}})$, and refit
both stages as in Section~\ref{FI_var}. Under (A1)--(A7),
\begin{equation}
\Prb^*\!\left[\sqrt n\{\widehat\zeta^*-\zeta(\widehat{\bm{\theta}})\}\le t\right]
\xrightarrow{p}\Phi_{\rm N}(t/\sqrt{V_{\rm FI}}),
\label{eq:FI_bootstrap_law}
\end{equation}
where $\Prb^*$ is conditional on the original data and $\Phi_{\rm N}$
is the standard normal distribution function. If the additional uniform
moment condition (A8) holds, then
$n\Var^*(\widehat\zeta^*)\xrightarrow{p}V_{\rm FI}$ and
$n\Var_0(\widehat\zeta)\to V_{\rm FI}$.
For $B=B_n\to\infty$ independent bootstrap replicates, the sample
variance in Section~\ref{FI_var} satisfies
$n\widehat V_{\rm BT}\xrightarrow{p}V_{\rm FI}$, and
$\widehat\zeta\pm z_{1-\alpha/2}\sqrt{\widehat V_{\rm BT}}$
has asymptotic coverage $1-\alpha$.
\end{theorem}

The bootstrap model has target $\zeta(\widehat{\bm{\theta}})$, which need not
equal the empirical FI average $\widehat\zeta$. Replacing this center
by $\widehat\zeta$ in~\eqref{eq:FI_bootstrap_law} can introduce a
first-order random shift. The bootstrap sample variance is invariant
to this choice of center. For a saturated finite categorical outcome
model, or when $h$ is an affine function of the sufficient statistics
of a canonical exponential family, exact score equations give
$\widehat\zeta=\zeta(\widehat{\bm{\theta}})$. Appendix C of Supporting Information proves the bootstrap
result and this distinction. Weak convergence alone does not imply
variance convergence; (A8) supplies the required moment control.

\subsection{Categorical data, Monte Carlo approximation, and contrasts}

For fixed finite categorical support, exact enumeration removes all
imputation Monte Carlo error. Appendix A of Supporting Information verifies (A8) under positive
cell probabilities, smooth identifiable models, nonsingular derivatives,
and exact global fits on compact parameter sets. This covers the type of
categorical analysis used in Sections~\ref{sec:simulation}
and~\ref{sec:application}, subject to these conditions. Convergence of
EM to an arbitrary stationary point alone is insufficient.

For continuous data, let $\widehat{\bm{\theta}}_m$ and
$\widehat\zeta_{{\rm FI},m}$ use $m=m_n$ proposal draws per incomplete
unit, fixed during optimization.
\begin{proposition}[Negligible imputation error]\label{prop:FI_MC}
Under (A1)--(A6) and Conditions (M1)--(M3) in Appendix C of Supporting Information,
\begin{align*}
\widehat{\bm{\theta}}_m-\widehat{\bm{\theta}}
&=\Op\{m^{-1}+(nm)^{-1/2}\}+\op(n^{-1/2}),\\
\widehat\zeta_{{\rm FI},m}-\widehat\zeta
&=\Op\{m^{-1}+(nm)^{-1/2}\}+\op(n^{-1/2}).
\end{align*}
Consequently, the sufficient rate condition
\begin{equation}
m_n/\sqrt n\longrightarrow\infty
\label{eq:FI_MC_rate}
\end{equation}
gives the same influence function and normal limit as exact evaluation.
\end{proposition}
The conditions require independent unit-specific proposal draws
conditionally on the observed data and pilot calculations,
$O(m^{-1})$ conditional bias and variance bounds, and local derivative
and optimization control. Proposal support alone does not establish
these bounds. Self-normalization can introduce $O(m^{-1})$ bias;
averaging reduces the centered imputation variance to $O((nm)^{-1})$,
but does not remove that bias. Thus $m\to\infty$ alone does not establish
root-$n$ equivalence, and a fixed choice such as $m=100$ is not justified
by an additional variance term alone. The bootstrap extension requires
the analogous conditional conditions; variance consistency additionally
requires the moment control specified in Appendix C of Supporting Information.

For a vector $H({\bm{Y}})$, define $m_H=\E_{\bm{\eta}}(H\mid O)$,
$D_{\bm{\phi}}=\E_0(\partial m_H/\partial{\bm{\phi}}\T)$, and
$D_{\bm{\theta}}=\E_0(\partial m_H/\partial{\bm{\theta}}\T)$. Then
\[
\psi_H=m_H-\zeta_H+
(D_{\bm{\phi}}-D_{\bm{\theta}} A_{\bm{\theta}}^{-1}C)A_{\bm{\phi}}^{-1}u_{\bm{\phi}}
+D_{\bm{\theta}} A_{\bm{\theta}}^{-1}u_{\bm{\theta}}.
\]
A smooth contrast $a(\zeta_H)$ has influence function
$\nabla a(\zeta_H)\T\psi_H$. In particular, for positive cell
probabilities,
\begin{equation}
\psi_{\log{\rm OR}}=\psi_{11}/p_{11}+\psi_{00}/p_{00}
-\psi_{10}/p_{10}-\psi_{01}/p_{01},
\label{eq:FI_logOR}
\end{equation}
and an interaction log odds ratio is a difference of such contrasts.
All results treat the graph as fixed and do not include uncertainty from
selecting a graph using the same data.
\end{theoryrevision}

\section{Simulation Study}\label{sec:simulation}

We consider a random vector $\bm{Y}=(Y_{1},Y_{2},Y_{3})$ of binary variables, each taking the value 0 or 1. Define the joint probability of these variables as $\pi_{ijk}=P(Y_{1}=i,Y_{2}=j,Y_{3}=k)$, modeled by a log-linear model with two-way interaction terms:
 \begin{align}\label{FI_sim_twoways}
 \log(\pi_{ijk}) = \lambda +\alpha_{i}+\beta_{j}+\gamma_{k}+(\alpha\beta)_{ij}+(\alpha\gamma)_{ik}+(\beta\gamma)_{jk},
 \end{align}
where $(\lambda,\alpha_{1},\beta_{1},\gamma_{1},(\alpha\beta)_{11},(\alpha\gamma)_{11})\!=\!(0, 0.125, 0.12, 0.15, -0.1, -0.1)$ with zero-sum restrictions, and $(\beta\gamma)_{11} \in \{0.12, 0.45\}$, representing weak and 
strong association between $Y_2$ and $Y_3$, respectively. We consider $\pi_{..1}=P(Y_{3}=1)$ and $\pi_{111}=P(Y_1 = 1, Y_{2}=1, Y_{3}=1)$ as the parameters of interest. 

We generate $M=1000$ Monte Carlo samples of size $n=500$ from $\text{Multinomial}(\bm{\pi}; n)$, where $\bm{\pi}\!=\!(\pi_{111},\hdots,\pi_{000})$. Since $Y_1$ is always observed, the response pattern set
$\mathcal{D}=\{111, 110, 101, 100\}$ is governed by the bivariate regular pattern graph for $(Y_2, Y_3)$ in
Figure~\ref{FI_4figs} (c). Given the full data $\bm{y}$, missingness is generated via:
\begin{align*}
P(\bm{\Delta}=111\mid\bm{Y}=\bm{y})&=\big[{1+O_{110}(y_{1},y_{2})+ O_{101}(y_{1},y_{3})+O_{100}(y_{1})\{O_{110}(y_{1},y_{2})+O_{101}(y_{1},y_{3})\}}\big]^{-1}\\
P(\bm{\Delta}=110\mid\bm{Y}=\bm{y})&=O_{110}(y_{1},y_{2})P(\bm{\Delta}=111\mid\bm{Y}=\bm{y})\\
P(\bm{\Delta}=101\mid\bm{Y}=\bm{y})&=O_{101}(y_{1},y_{3})P(\bm{\Delta}=111\mid\bm{Y}=\bm{y})\\
P(\bm{\Delta}=100\mid\bm{Y}=\bm{y})&=O_{100}(y_1)\{O_{110}(y_1,y_2)+O_{101}(y_{1}, y_3)\}P(\bm{\Delta}=111\mid\bm{Y}=\bm{y}),
\end{align*}
where the conditional odds are defined as follows: $O_{110}(y_{1},y_{2}) = \exp(\beta_{110,0}+\beta_{110,1}y_{1}+\beta_{110,2}y_{2})$, $O_{101}(y_{1},y_{3}) = \exp(\beta_{101,0}+\beta_{101,1}y_{1}+\beta_{101,2}y_{3})$, and $O_{100}(y_{1}) = \exp(\beta_{100,0}+\beta_{100,1}y_{1})$. The parameter configurations and the Monte Carlo (MC) proportions for each missing pattern are summarized in Table~\ref{tab:scenario_S0}.
\begin{table}[ht]
\centering
\renewcommand{\arraystretch}{1.2}
\scalebox{0.8}{
\begin{tabular}{cccccccccccccc}
  \hline
Scenario & $(\beta\gamma)_{11}$ & $\beta_{110,0}$ & $\beta_{110,1}$ & $\beta_{110,2}$ & $\beta_{101,0}$ & $\beta_{101,1}$ & $\beta_{101,2}$ & $\beta_{100,0}$ & $\beta_{100,1}$ & $\hat{p}_{111}$ & $\hat{p}_{110}$ & $\hat{p}_{101}$ & $\hat{p}_{100}$ \\
  \hline
1 (Hi/Wk) & 0.12 & $-$0.3 & 0.7 & $-$0.3 & $-$0.3 & $-$0.8 & 0.3 & $-$1.5 & 0.5 & 0.328 & 0.308 & 0.206 & 0.159 \\
2 (Hi/St) & 0.45 & $-$0.3 & 0.7 & $-$0.3 & $-$0.3 & $-$0.8 & 0.3 & $-$1.5 & 0.5 & 0.328 & 0.308 & 0.208 & 0.156 \\
3 (Lo/Wk) & 0.12 & 1.8 & 0.7 & $-$0.3 & 1.8 & $-$0.8 & 0.3 & $-$0.5 & 0.5 & 0.041 & 0.310 & 0.221 & 0.429 \\
4 (Lo/St) & 0.45 & 1.8 & 0.7 & $-$0.3 & 1.8 & $-$0.8 & 0.3 & $-$0.5 & 0.5 & 0.040 & 0.314 & 0.221 & 0.425 \\
  \hline
\end{tabular}}
\caption{Parameter configuration for the response mechanism model and MC proportions for each missing pattern. Hi/Lo: High/Low full response rate (FR$\approx$32\%/4\%). Wk/St: Weak/Strong association ($(\beta\gamma)_{11}=0.12/0.45$).}
\label{tab:scenario_S0}
\end{table}

We further examine the estimator's sensitivity to the working model's parametric complexity by comparing a parsimonious (two-way interaction) specification against the fully saturated specification. That is, we also consider the saturated working model; the corresponding estimator is denoted by $\hat{\zeta}_{\rm FI.ST}$. The two estimators, $\hat{\zeta}_{\rm FI}$ and $\hat{\zeta}_{\rm FI.ST}$, differ only in the M-step of the EM algorithm. Under the two-way interaction model for the three categorical variables, we can maximize $Q(\bm{\theta}\mid\bm{\theta}^{(t)})$ at the $t$-th iteration using the iterative proportional fitting algorithm. Note that model (\ref{FI_sim_twoways}) implies that $n_{ijk}^{(t+1)}$ should satisfy
\begin{align*}
1=n_{ij.}^{(t)}/n_{ij.}^{(t+1)}=n_{i.k}^{(t)}/n_{i.k}^{(t+1)}=n_{.jk}^{(t)}/n_{.jk}^{(t+1)},
\end{align*}
where $n_{ijk}^{(t)}$ is the estimated frequency for category $(i,j,k)$ at the $t$-th iteration. Then, the iterative proportional fitting algorithm can be implemented by repeating the following updates, starting from initial guesses $\hat{c}_{ijk}^{[0]}$: $$\hat{c}_{ijk}^{[3s+1]} = \frac{n_{ij.}^{(t)}}{\hat{c}_{ij.}^{[3s]}}\hat{c}_{ijk}^{[3s]}\rightarrow \hat{c}_{ijk}^{[3s+2]} = \frac{n_{i.k}^{(t)}}{\hat{c}_{i.k}^{[3s+1]}}\hat{c}_{ijk}^{[3s+1]}\rightarrow\hat{c}_{ijk}^{[3(s+1)]} = \frac{n_{.jk}^{(t)}}{\hat{c}_{.jk}^{[3s+2]}}\hat{c}_{ijk}^{[3s+2]}.$$
Under the saturated model, there is no restriction on marginals, since the model includes all possible interaction terms. Thus, $n_{ijk}^{(t+1)} =n_{ijk}^{(t)}$ \citep{agresti2012categorical,christensen2006log,von2012log}.

We evaluate the performance of the different estimation methods described in Sections~\ref{sec:setup} and~\ref{sec:proposed}. The estimates obtained using these methods are denoted as $\hat{\zeta}_{t}^{(m)}$, where $t\in \{\text{IPW, RA, AIPW, FI, FI.ST}\}$ corresponds to the method used at the $m$-th simulation iteration. To assess their performance, we calculate the MC bias (${\rm BIAS}_{t}$), standard deviation (${\rm SD}_{t}$), and root-mean-squared error (${\rm RMSE}_{t}$). BIAS$_{t}$ is computed as $M^{-1}\sum_{m=1}^{M}(\hat{\zeta}_{t}^{(m)}-\zeta{_0})$, the ${\rm SD}_{t}$ as $\sqrt{M^{-1}\sum_{m=1}^{M}(\hat{\zeta}_{t}^{(m)}-\bar{\zeta}_{{\rm MC},t})^{2}}$ with $\bar{\zeta}_{{\rm MC},t}=M^{-1}\sum_{m=1}^{M}\hat{\zeta}_{t}^{(m)}$, and the ${\rm RMSE}_{t}$ as $\sqrt{M^{-1}\sum_{m=1}^{M}(\hat{\zeta}_{t}^{(m)}-{\zeta}_{0})^{2}}$. The resulting performance metrics are presented in Table~\ref{tab:S0_results}.

\begin{table}[ht]
\centering
\renewcommand{\arraystretch}{1.2}
\caption{Simulation results under the true two-way interaction model, $M=1000$, $n=500$. 
All values are multiplied by 100. Hi/Lo: High/Low full-response rate (FR $\approx$32\%/4\%). Wk/St: Weak/Strong association ($(\beta\gamma)_{11}=0.12/0.45$).}
\label{tab:S0_results}
\scalebox{0.9}{
\begin{tabular}{ccccccccc}
\toprule
 &  & \multicolumn{3}{c}{$\pi_{..1}$} & & \multicolumn{3}{c}{$\pi_{111} $} \\
\cmidrule(lr){3-5} \cmidrule(lr){7-9}
Scenario & Method & BIAS & SD & RMSE & & BIAS & SD & RMSE \\
\midrule
1 (Hi/Wk) & IPW   & $-$0.133 & 3.207 & 3.209 & & $-$0.029 & 2.432 & 2.432 \\
          & RA    & $-$0.139 & 3.227 & 3.230 & & $-$0.012 & 2.229 & 2.229 \\
          & AIPW  & $-$0.139 & 3.227 & 3.230 & & $-$0.012 & 2.229 & 2.229 \\
          & FI    & $-$0.132 & 3.200 & 3.203 & &    0.004 & 2.045 & 2.045 \\
          & {FI.ST} & ${-0.133}$ & ${3.210}$ & ${3.213}$ & & ${-0.013}$ & ${2.214}$ & ${2.214}$ \\
\midrule
2 (Hi/St) & IPW   & $-$0.086 & 3.076 & 3.078 & & $-$0.005 & 2.662 & 2.662 \\
          & RA    & $-$0.102 & 3.075 & 3.077 & & $-$0.027 & 2.374 & 2.374 \\
          & AIPW  & $-$0.102 & 3.075 & 3.077 & & $-$0.027 & 2.374 & 2.374 \\
          & FI    & $-$0.090 & 3.055 & 3.057 & &    0.004 & 2.235 & 2.235 \\
          & {FI.ST} & ${-0.092}$ & ${3.058}$ & ${3.059}$ & & ${-0.027}$ & ${2.377}$ & ${2.377}$ \\
\midrule
3 (Lo/Wk) & IPW & 0.012 & 7.863 & 7.863 & & 0.245 & 7.623 & 7.627\\
 & RA & $-$0.061 & 8.297 & 8.297 & &0.244 & 6.354 & 6.359\\
 & AIPW & $-$0.061 & 8.297 & 8.297 & &0.244 & 6.354 & 6.359\\
 & FI & 0.066 & 7.501 & 7.501 & &0.063 & 4.548 & 4.548\\
 & FI.ST & 0.024 & 7.632 & 7.632 & &0.276 & 5.707 & 5.714\\
\midrule
4 (Lo/St) & IPW & $-$0.238 & 7.653 & 7.657 & &0.091 & 7.281 & 7.281\\
 & RA &$-$0.117 & 7.898 & 7.899 & &0.021 & 5.454 & 5.454\\
 & AIPW & $-$0.117 & 7.898 & 7.899 & &0.021 & 5.454 & 5.454\\
 & FI & $-$0.060 & 7.089 & 7.089 & &$-$0.167 & 4.128 & 4.132\\
 & FI.ST & $-$0.012 & 7.185 & 7.185& & $-$0.042 & 4.942 & 4.942\\
\bottomrule
\end{tabular}}
\end{table}

\begin{figure}
    \centering
    \includegraphics[width=1\linewidth]{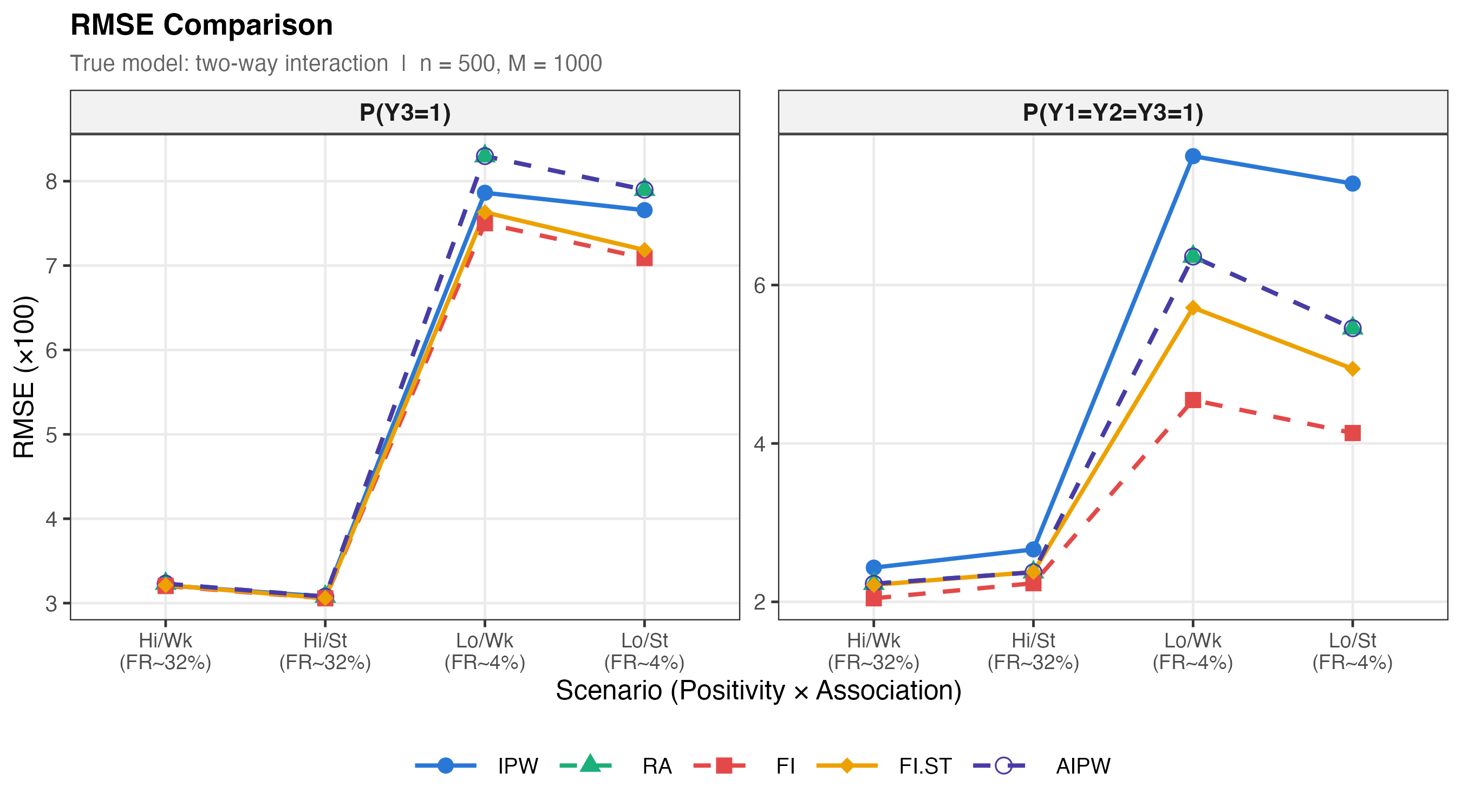}
    \caption{RMSE (\%) of each estimator under the true two-way interaction model for $\pi_{..1}=P(Y_3=1)$ and $\pi_{111}=P(Y_1=Y_2=Y_3=1)$.}
    \label{fig:rmse_S0}
\end{figure}

Across all scenarios, the bias of each method has a negligible impact on the RMSE, so we focus on interpreting the results in terms of SD and RMSE, which reflect the efficiency of each method rather than its accuracy. 

With a high full-response rate (FR $\approx$ 32\%), the RMSE values of all methods are very close to each other for $\pi_{..1}$. For the joint parameter $\pi_{111}$, the differences between methods are slightly larger but still minor. In this high-response setting, model complexity and the strength of association do not lead to a meaningful efficiency gain, since there are enough fully observed cases for all methods to be stable.

RA and AIPW yield identical values throughout Table~\ref{tab:S0_results}. This is an algebraic identity rather than a coincidence. Because all variables are categorical, the RA imputation reduces to stratum-specific sample means, and the score equations of the maximum likelihood fit of the logistic odds models force the AIPW correction term to sum to zero
within these strata, so that AIPW reduces exactly to RA. This identity breaks down only when a covariate stratum used in the donor-based imputation contains no fully observed cases, which occurs in about 6--7\% of the generated data sets in the low-response scenarios. To keep the bias, SD, and RMSE of all methods comparable on the same $M=1000$
replicates, we generated additional draws and retained only those in which every stratum was populated; the low-response entries in Table~\ref{tab:S0_results} therefore describe performance conditional on this event rather than averaged over it. A practitioner working with a single data set cannot redraw in this way and should instead merge or
re-parent sparse patterns, as recommended in Section~\ref{sec:application}.

When the full-response rate drops (FR $\approx$ 4\%), the efficiency gap between methods becomes much clearer (Figure~\ref{fig:rmse_S0}). IPW and AIPW lose efficiency, with RMSE for $\pi_{111}$ rising to 5.454--7.627. In particular, the AIPW estimator is more variable than the plain IPW estimator for $\pi_{\cdot\cdot1}$. With a full-response rate near 4\%, the estimated weights $1/\hat\pi_{\bm{1}}(\by)$ are large and unstable, and the augmentation term -- which improves efficiency when the weights are accurate -- can amplify this instability in finite samples. This behavior is consistent with the well-documented sensitivity of weighting-based estimators under weak positivity, and it is precisely the regime in which the likelihood-based FI estimator, which does not weight by $1/\hat\pi_{\bm{1}}$, is most advantageous. FI is the most efficient method in this low-response setting, with the smallest RMSE for both $\pi_{..1}$ (7.089--7.501) and $\pi_{111}$ (4.132--4.548). For the joint parameter $\pi_{111}$, the more complex methods (FI, FI.ST) gain a clearer efficiency advantage over the simpler ones (IPW, RA) once information on fully observed triples becomes scarce. FI.ST is slightly less efficient than FI but still outperforms IPW, RA, and AIPW in every low-response scenario. These results suggest that the proposed FI estimator effectively borrows information across response patterns by jointly estimating the complete-data model, leading to improved efficiency when fully observed cases are scarce.

We also examine the accuracy of the bootstrap variance estimator (BT), presented in Section \ref{FI_var}, for $\hat{\zeta}_{\rm FI.ST}$ in the scenario with a high full-response rate and weak association. To measure the precision of BT, we calculate the relative bias (RB) with respect to the MC variance, given by ${\rm RB} = ({\rm Var}_{{\rm BT}}-{\rm Var}_{{\rm MC}})/{{\rm Var}_{{\rm MC}}},$ where 
${\rm Var}_{{\rm BT}}=M^{-1}\sum_{m=1}^{M}\widehat{\rm Var}_{{\rm BT}}^{(m)}$ and 
${\rm Var}_{{\rm MC}}={M^{-1}\sum_{m=1}^{M}(\hat{\zeta}_{t}^{(m)}-\bar{\zeta}_{{\rm MC},t})^{2}}$. The empirical coverage probability (ECP) of a 95\% confidence interval constructed using the bootstrap variance estimator is also calculated as 
$${\rm ECP}=M^{-1}\sum_{m=1}^{M}I(\zeta_{0}\in {\rm CI}^{(m)}),$$
where ${\rm CI}^{(m)}=\hat{\zeta}^{(m)}\pm 1.96\sqrt{\widehat{\rm Var}_{{\rm BT}}^{(m)}}$.

\begin{figure}
    \centering
    \includegraphics[width=0.7\linewidth]{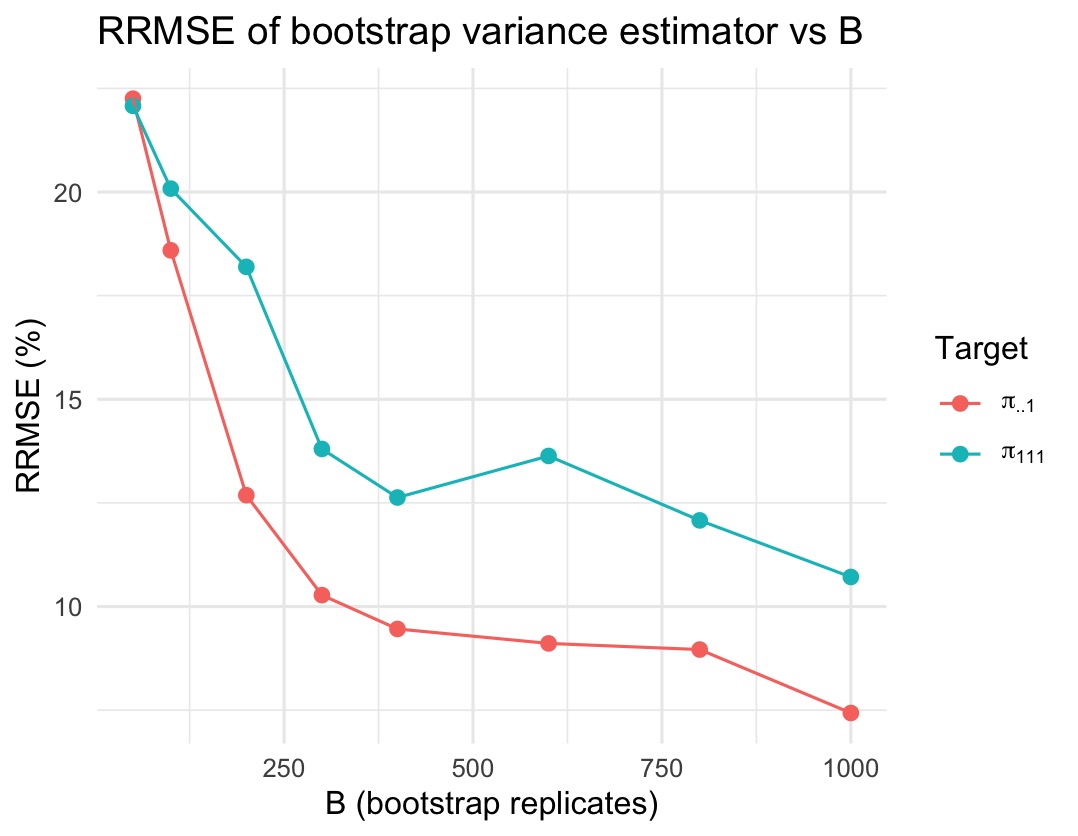}
    \caption{RRMSE (\%) of the bootstrap variance estimator versus the number of bootstrap replicates $B$, for $\pi_{..1}$ and $\pi_{111}$. }
    \label{fig:RRMSE_B}
\end{figure}

Figure~\ref{fig:RRMSE_B} and Table~\ref{tab:diag_by_B} summarize the performance of the proposed parametric bootstrap variance estimator as the number of bootstrap replicates, $B$, increases. As expected, the relative RMSE (RRMSE) decreases steadily with $B$, indicating increasing stability of the bootstrap variance estimator. Consistent with this pattern, the relative bias (RB) remains small for both $\pi_{..1}$ and $\pi_{111}$, with the absolute RB generally below 9\% and below 1\% when $B=1000$. Moreover, the ECP remains close to the nominal 95\% level across all values of $B$. \theorychange{These results support the use of $B=1000$ in this baseline experiment. They are a finite-sample stability check; the consistency result in Theorem~\ref{thm:FI_bootstrap} uses $B_n\to\infty$.}

\begin{table}[ht]
\centering
\renewcommand{\arraystretch}{1.3}
\caption{Relative bias (RB, \%) of the bootstrap variance estimator relative to the Monte Carlo variance, and empirical coverage probability (ECP), by $B$ for $\pi_{..1}$ and $\pi_{111}$ under the baseline case (Hi/Wk). The MC and BT columns report variances multiplied by $10^{3}$.}
\label{tab:diag_by_B}
\begin{tabular}{ccccc|cccc}
  \hline
\multirow{2}{*}{$B$} & \multicolumn{4}{c}{$\pi_{..1}$} & \multicolumn{4}{c}{$\pi_{111}$} \\ \cline{2-9}
 & MC & BT & RB & ECP & MC & BT & RB & ECP \\
  \hline
  50 & 0.992 & 1.033 & 4.086 & 0.952 & 0.550 & 0.539 & -2.077 & 0.940 \\ 
  100 & 0.943 & 1.024 & 8.584 & 0.942 & 0.500 & 0.532 & 6.339 & 0.954 \\ 
  200 & 0.989 & 1.019 & 3.116 & 0.948 & 0.496 & 0.540 & 8.734 & 0.960 \\ 
  300 & 1.042 & 1.024 & -1.657 & 0.954 & 0.526 & 0.542 & 3.120 & 0.954 \\ 
  400 & 1.035 & 1.029 & -0.576 & 0.950 & 0.557 & 0.542 & -2.812 & 0.944 \\ 
  600 & 0.983 & 1.023 & 4.047 & 0.964 & 0.518 & 0.543 & 4.760 & 0.948 \\ 
  800 & 1.076 & 1.024 & -4.849 & 0.936 & 0.528 & 0.541 & 2.404 & 0.948 \\ 
  1000 & 1.017 & 1.024 & 0.681 & 0.950 & 0.535 & 0.539 & 0.736 & 0.946 \\
 
   \hline
\end{tabular}
\end{table}

\section{Implementation and Data Analysis}\label{sec:application}

In this section, we turn from methodology to practice. Section~\ref{sec:guidelines} summarizes a step-by-step workflow for applying the proposed method, from specifying candidate pattern graphs to fitting the odds and outcome models and
estimating variances. Section~\ref{sec:mi_application} then illustrates this workflow with an analysis of the associations of opioid administration and hypotension with mortality in patients with myocardial infarction, where two
key clinical variables are subject to non-monotone missingness.

\subsection{Practical guidelines} \label{sec:guidelines}

For practitioners we summarize the workflow as follows.

\begin{enumerate}
\item \textbf{Enumerate the patterns.} List the observed response patterns with their counts,
and verify that the complete pattern \(\mathbf{1} = (1,\ldots,1)\) occurs with adequate
frequency; the positivity condition of Section~\ref{sec:setup} requires every pattern--parent comparison
to be populated.

\item \textbf{Specify candidate graphs.} Begin with the star graph (PG1), in which every incomplete
pattern has the single parent \(\mathbf{1}\), and add or redirect edges suggested by
subject-matter structure (for example, temporal ordering of visits in longitudinal studies),
respecting the regularity conditions that \(\mathbf{1}\) is the unique source and that parents
strictly dominate children.

\item \theorychange{\textbf{Fit the odds models.} There is one logistic
odds model per non-source pattern, hence $|\mathcal D|-1$ models, regardless
of the number of edges. Compare each $\bm\delta$ with its pooled parent
set using~\eqref{logistic}. Check for separation, sparse comparisons, and
extreme fitted probabilities. The theory assumes a fixed graph; changes
prompted by sparsity should be reported and assessed through explicitly
specified alternative graphs.}

\item \theorychange{\textbf{Estimate the outcome model.} Run the EM/FI
algorithm of Section~\ref{sec:proposed} under each graph. For categorical
outcomes, enumerate the missing-value support. For continuous outcomes,
check proposal support, weight stability, and sensitivity to the
imputation count; support alone does not justify negligible imputation
error. Check convergence to an appropriate likelihood solution.}

\item \theorychange{\textbf{Compare graphs as a sensitivity analysis.}
Compare held-out observed-data log-likelihoods and report estimates under
all plausible graphs. Likelihood-based AIC/BIC rankings may be reported
as descriptive diagnostics; their usual joint-MLE calibration is not
established here for the two-step fit. Substantive conclusions should be
assessed across graphs because the identifying restrictions are only
partially testable.}

\item \textbf{Estimate variances.} Use the parametric bootstrap of Section~\ref{FI_var}
\theorychange{with a sufficiently large $B$, re-estimating both $\bm\phi$
and $\bm\theta$ in every replicate. A value such as $B=1000$ is a practical
starting point; assess Monte Carlo stability. Report inference separately
for each fixed graph, since the theory does not account for graph
selection using the same data.}
\end{enumerate}
\subsection{Application to myocardial infarction complications data}\label{sec:mi_application}

We illustrate the workflow of Section~\ref{sec:guidelines} on a clinical data set concerning
opioid administration and mortality following myocardial infarction.

Myocardial infarction (MI), commonly known as a heart attack, occurs when blood flow to the heart muscle is blocked, leading to severe cardiac complications and potentially lethal outcomes. In emergency settings, opioids are frequently administered to provide chest pain relief. However, prior studies have reported cardiovascular side effects of opioid administration, including vasodilation, bradycardia, and orthostatic intolerance \citep{krantz2021cardiovascular, chen2015cardiac, hristovska2023orthostatic}. These adverse effects may contribute to reductions in blood pressure and may be particularly concerning in patients who already present with low baseline blood pressure. These findings motivated us to investigate whether opioid use in patients with low baseline blood pressure is associated with an increased risk of fatal outcomes following MI.

To answer this research question, we used the clinical MI data of \citet{myocardial_infarction_complications_579}, publicly available from the UCI Machine Learning Repository. The dataset contains 123 variables, among which we focused on \texttt{LET\_IS} (lethal outcome), \texttt{NA\_KB} (opioid use in emergency care), and \texttt{S\_AD\_KBRIG} (systolic blood pressure measured by the Emergency Cardiology Team). The variable \texttt{LET\_IS} was originally
categorical, indicating survival or one of seven specific causes of death. Of the full analytic cohort ($n=1{,}700$), 271 patients (15.9\%) died; because the objective of this study was to investigate overall mortality following myocardial infarction, and because the resulting cause-specific death counts (ranging from 12 to 110) were already too sparse -- even
using the complete available sample -- to support reliable cause-specific modeling within our framework, \texttt{LET\_IS} was recoded as a binary outcome indicating death (1) versus survival (0). Similarly, the variable \texttt{S\_AD\_KBRIG} was dichotomized at 90~mmHg ($1 = \text{SBP} \leq 90$~mmHg, $0 = \text{SBP} > 90$~mmHg), following the standard clinical
definition of hypotension \citep{eastridge2007hypotension}; this threshold was adopted for consistency with the broader clinical literature and because our aim was to evaluate the opioid--mortality association specifically among patients meeting the established clinical definition of hypotension, rather than to identify a data-driven optimal
cutoff.

\begin{figure}
    \centering    \includegraphics[width=1\linewidth]{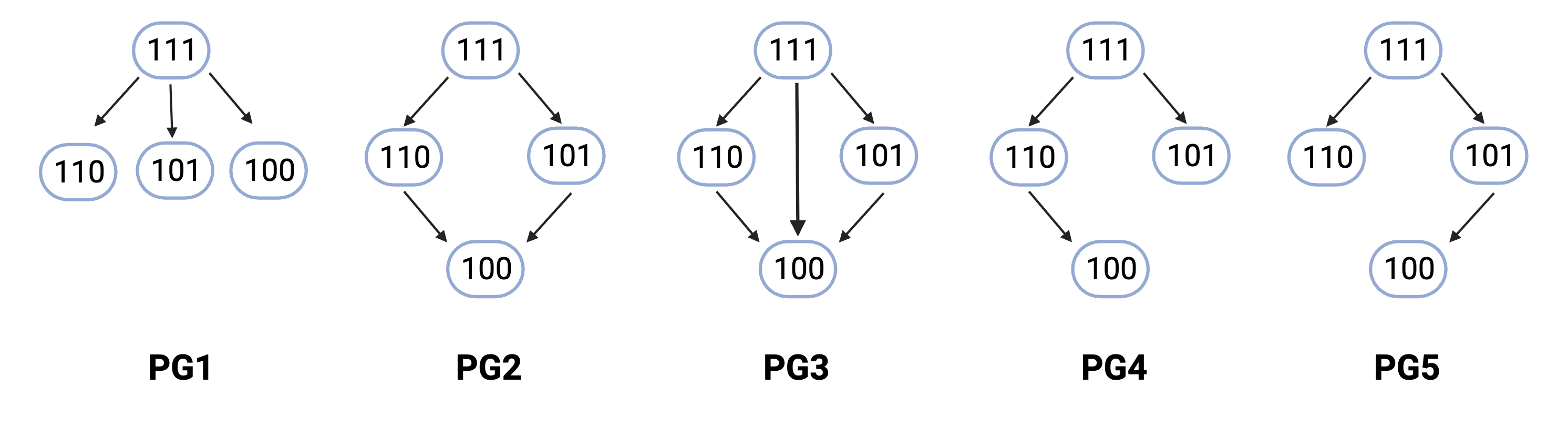}
    \caption{Candidate pattern graphs for the missing patterns defined by (\texttt{LET\_IS}, \texttt{NA\_KB}, \texttt{S\_AD\_KBRIG}).}
    \label{fig:data_pgs}
\end{figure}
\begin{figure}
    \centering
    \includegraphics[width=1\linewidth]{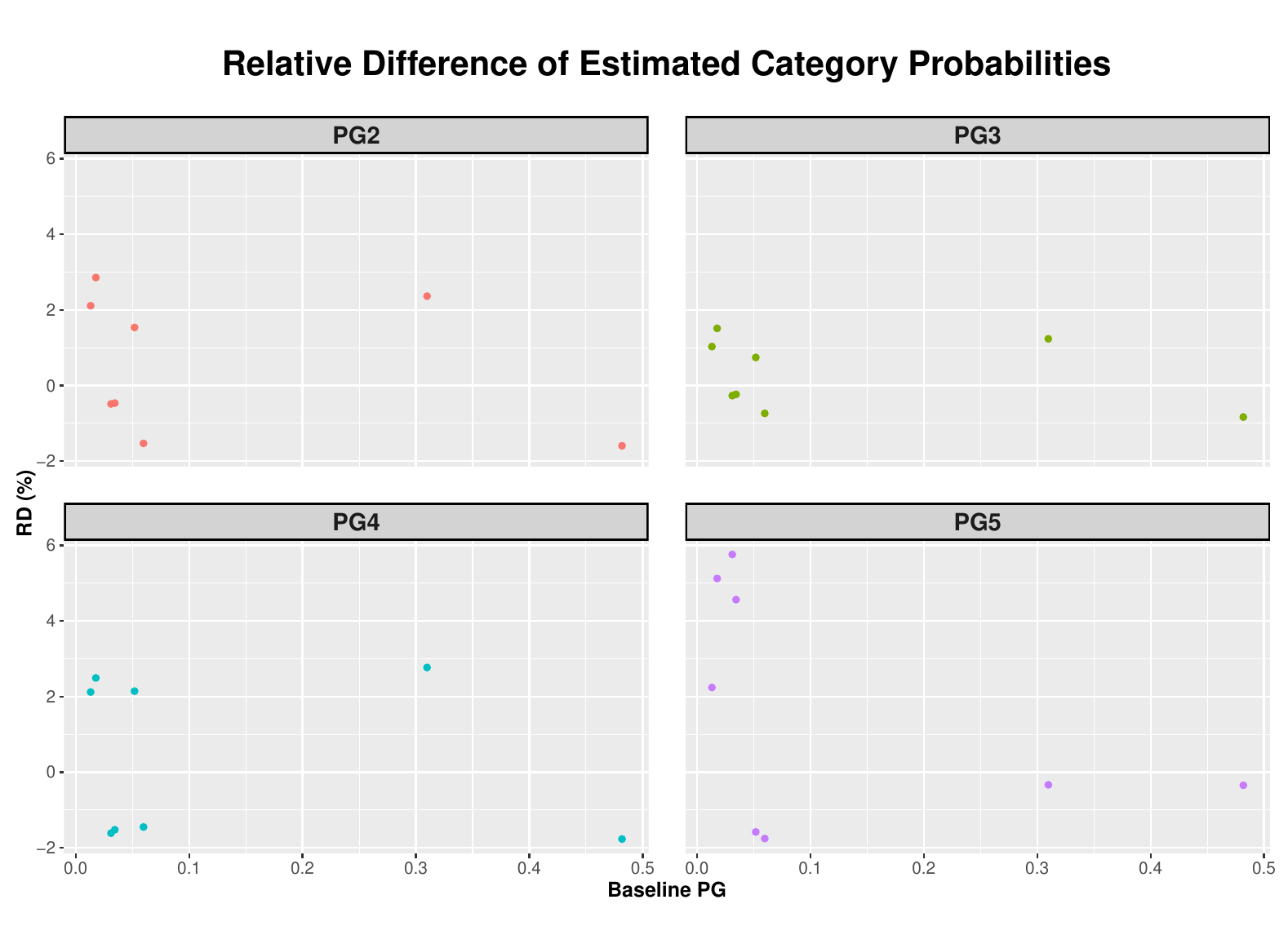}
    \caption{Relative differences (RD, \%) in estimated category probabilities for candidate pattern graphs.}
    \label{fig:relativedifference}
\end{figure}

The analysis data set comprises \(n = 1{,}700\) patients. The proportions of missing values were 0 for \texttt{LET\_IS}, 0.386 for \texttt{NA\_KB}, and 0.633 for \texttt{S\_AD\_KBRIG}. Writing the response pattern as a triplet (\texttt{LET\_IS}, \texttt{NA\_KB}, \texttt{S\_AD\_KBRIG}), four patterns occur: \((1,1,1)\) with 545 patients (32.1\%), \((1,1,0)\) with 498 (29.3\%), \((1,0,1)\) with 79 (4.6\%), and \((1,0,0)\) with 578 (34.0\%). Because \texttt{LET\_IS} is fully observed, the missingness structure is driven by (\texttt{NA\_KB}, \texttt{S\_AD\_KBRIG}), and all four patterns retain the outcome variable. A logistic regression model was fitted with the missingness indicator of \texttt{NA\_KB} as the response and \texttt{LET\_IS} as the predictor. The odds of missingness were significantly lower among patients with lethal outcomes (OR = 0.685, 95\% CI 0.517--0.901; p = 0.007), indicating that the probability of missingness depended on the observed outcome. Therefore, the missing completely at random (MCAR) assumption was rejected. For the missing patterns defined by (\texttt{NA\_KB}, \texttt{S\_AD\_KBRIG}), we considered five pattern graph models in Figure~\ref{fig:data_pgs}, and for the data model specification we conservatively adopted a saturated log-linear model including all main effects, two-way interactions, and the three-way interaction among (\texttt{LET\_IS}, \texttt{NA\_KB}, \texttt{S\_AD\_KBRIG}) to avoid potential bias from model misspecification. Under this specification, using the imputation method of Section~\ref{sec:proposed}, PG1 achieved the largest observed log-likelihood among the five candidate graphs ($-2362.894$, versus $-2371.002$, 
$-2367.177$, $-2369.586$, and $-2377.389$ for PG2--PG5, respectively). As shown in Figure~\ref{fig:relativedifference}, the category-level relative differences are modest for all candidate graphs, with PG2--PG4 yielding estimates very similar to PG1.

Because the candidate graphs encode identification restrictions that are only partially
testable from the observed data, we regard the comparison across PG1--PG5 primarily as a
sensitivity analysis for the identification assumption rather than as model selection. Two
remarks support this reading. First, although the five graphs differ in the parent pattern
assigned to $(1,0,0)$ (the fully observed pattern, a pooled combination of patterns, or a single incompletely observed pattern), the fitted odds model for $(1,0,0)$ depends on \texttt{LET\_IS} alone in every graph, and the models for $(1,1,0)$, $(1,0,1)$, and the outcome distribution are common to all five graphs; consequently PG1--PG5 have an identical number of free parameters ($k=15$), and AIC and BIC are monotone transformations of the observed log-likelihoods that yield the same ranking, which favors PG1. Second, and more importantly, Figure~\ref{fig:relativedifference} shows that the estimated category probabilities agree within about 2\% across PG1--PG4, while PG5 deviates somewhat more (up to about 6\% in a few categories) but still by a modest margin; the substantive conclusions are therefore robust to the choice among the plausible graphs. Across 100 stratified 80/20 train--test splits, all five graphs attain nearly identical out-of-sample observed log-likelihoods (ranging from $-752.3$ to $-752.5$); bootstrap analysis ($B=500$) shows PG1, PG2, and PG4 to be the most stable for the target interaction odds ratio (bootstrap SD of the log-OR: $0.67$, $0.67$, and $0.70$, respectively), with PG3 and PG5 noticeably less stable ($1.01$ for both). PG1 attains the highest full-data observed log-likelihood and has the simplest structure among the candidate graphs, with every incomplete pattern parented directly
by the fully observed pattern; we therefore retain it as the primary specification on grounds of fit, stability, and parsimony. The derived odds ratios for PG2--PG5, reported in Appendix E of Supporting Information, are consistent with the main-text estimates across every graph considered.

\begin{theoryrevision}
For a fixed graph, the odds ratios are smooth functions of
the FI estimates of the joint cell probabilities. Their joint asymptotic
distribution follows from~\eqref{eq:FI_logOR} when the relevant cell
probabilities are positive and the limiting contrast variance is
nonzero. With the saturated outcome model and exact score solutions,
the FI cell estimates equal the fitted model cell probabilities.
The inferential guarantees treat each graph as fixed; they do not cover
uncertainty introduced by choosing PG1 after comparing graphs.
\end{theoryrevision}

In our analyses, all reported odds ratios are obtained from the proposed FI estimator, which incorporates the incompletely observed patterns through the fractional weights described in Section \ref{sec:proposed}; as Table \ref{tab:full_stratified} shows, this corresponds to an effective sample of about 165 patients in the hypotensive stratum, rather than only the 57 patients with complete data. Hypotension was a strong predictor of mortality both among patients not receiving opioids (OR = 4.31, 95\% CI 1.49--12.44; p = 0.007) and, more strongly, among those receiving opioids (OR = 8.70, 95\% CI 4.23--17.91; p $<$ 0.001). Opioid use itself was not significantly
associated with mortality, either among non-hypotensive patients (OR = 0.72, 95\% CI 0.48--1.08) or among hypotensive patients (OR = 1.45, 95\% CI 0.44--4.73). The interaction between opioid use and hypotension status was also not significant (OR = 2.02, 95\% CI
0.54--7.51). The analysis therefore identifies hypotension as a robust and dominant risk factor for mortality among these myocardial infarction patients, regardless of opioid status, while
opioid administration shows no significant association with mortality in either stratum.

\begin{table}[h]
\centering
\caption{FI-based effective sample sizes by opioid use and hypotension status
(fractional counts reflecting the incompletely observed patterns; total $n=1{,}700$).}
\label{tab:full_stratified}
\begin{tabular}{llccc}
\toprule
Opioid & Hypotension & Survived & Died & Total \\
\midrule
No opioid        & Non-hypotensive & 528.5 & 90.6  & 619.1  \\
Received opioid  & Non-hypotensive & 815.8 & 100.1 & 916.0  \\
No opioid        & Hypotensive     & 30.9  & 22.8  & 53.7   \\
Received opioid  & Hypotensive     & 53.7  & 57.4  & 111.2  \\
\midrule
Total            &                 & 1429.0 & 271.0 & 1700.0 \\
\bottomrule
\end{tabular}
\end{table}

\section{Concluding Remarks}\label{sec:conclusion}

We have proposed a fractional-imputation-based estimation procedure for non-monotone and nonignorable missing data within a regular pattern graph. The proposed method gains efficiency over the existing methods of \citet{chen2022pattern}, particularly when fully observed cases are scarce, as confirmed by the simulation study. In the myocardial infarction application, incorporating the incompletely observed patterns roughly tripled the effective sample size in the hypotensive stratum. The choice of the pattern graph is itself a model selection problem. Conceptually, one could use cross-validation and choose the graph with the smallest mean squared prediction error, but such an approach is computationally expensive. An alternative approach based on information criteria could also be developed; we leave this as a topic for future research.

\begin{theoryrevision}
The consistency result for $\hat\zeta_{\rm FI}$ assumes correct
specification of the pattern graph, the odds models, and the outcome
model. These are sufficient conditions for the result proved here, not
a claim that every component must be correctly specified for every
possible target or every likelihood-based procedure. The first-stage odds estimates can
remain consistent under outcome-model misspecification, but this alone
does not establish consistency of FI. We do not claim double robustness
for the proposed estimator. In light of the robustness results in the pattern-graph
literature \citep{chen2022pattern}, augmenting FI with inverse-probability
weighting is a possible direction for further work. Our variance results also assume a fixed graph; inference after
data-dependent graph selection requires additional analysis.
\end{theoryrevision}

\section*{Funding}
This work is supported by USDA Natural Resources Conservation Service [017301-00001 to E.B.], and the U.S. Department of Agriculture’s National Resources Inventory, Cooperative Agreement NR203A750023C006, Great Rivers CESU 68-3A75-18-504. The research of Kim was partially supported by a grant from US National Science Foundation (2242820), a grant from the Iowa Agriculture and Home Economics Experiment Station (Project No. IOW05806).

\section*{Use of AI Tools}
   Claude (Anthropic) was used for language editing and for assistance with checking code, references, and LaTeX formatting after the authors wrote the initial draft. The authors take full responsibility for the content. 

\section*{Conflict of Interest}
The authors declare no potential conflicts of interest.

\section*{Data Availability Statement}
The data used in this paper can be found at \url{https://archive.ics.uci.edu/dataset/579/myocardial+infarction+complications} (DOI: 10.24432/C53P5M)~\citep{myocardial_infarction_complications_579}. 

\section*{Code Availability Statement}
The R codes  relevant to the simulation study and data analysis in Sections \ref{sec:simulation} and \ref{sec:application} are available at \url{https://github.com/yhcho111/Pattern-Graph}.

\bibliographystyle{unsrtnat}   
\bibliography{ref}
\section*{Supporting Information}
\theorychange{Appendix A states the regularity conditions and a finite categorical result. Appendix B proves Proposition~\ref{prop:eta_consistency}, Theorem~\ref{thm:FI_normality}, and sandwich consistency. Appendix C proves Theorem~\ref{thm:FI_bootstrap} and Proposition~\ref{prop:FI_MC}, including the additional Monte Carlo conditions. Appendices D and E provide the continuous-data simulation and graph-sensitivity results, respectively.}

\end{document}